\documentclass[fleqn,usenatbib]{mnras}

\usepackage{newtxtext,newtxmath}
\usepackage[T1]{fontenc}
\usepackage{ae,aecompl}

\usepackage{graphicx}   % Including figure files
\usepackage{amsmath}    % Advanced maths commands
\usepackage{amssymb}    % Extra maths symbols
\usepackage{todonotes}
\graphicspath{{img/}}   % Set image path

\newcommand{\satellite}[1]{\textit{#1}}  % Satellite names italic
\newcommand{\code}[1]{\textsc{#1}}       % Code names small caps
\newcommand{\Sun}[0]{\ensuremath{_{\odot}}}
\renewcommand{\deg}{\ensuremath{^{\circ}}}

\title[Cygnus~A merger simulations]{Simulations of the merging cluster of galaxies Cygnus~A}

\author[T. L. R. Halbesma]{
T. L. R. Halbesma,$^{1,2}$\thanks{E-mail: \href{mailto:halbesma@MPA-Garching.MPG.DE}{halbesma@MPA-Garching.MPG.DE}}
J. M. F. Donnert,$^{3,4}$
M. N. de Vries,$^{1,5}$
M. W. Wise$^{1,5}$%\thanks{E-mail: \href{mailto:wise@astron.nl}{wise@astron.nl}}
\\
$^{1}$Astronomical Institute ``Anton Pannekoek'', University of Amsterdam, Science Park 904, 1098 XH Amsterdam, The Netherlands \\
$^{2}$Max Planck Institute for Astrophysics, Karl-Schwarzschild-Str. 1, D-85741 Garching, Germany\\
$^{3}$INAF Istituto di Radioastronomia, via P. Gobetti 101, I-40129 Bologna, Italy \\
$^{4}$University of Minnesota, School of Physics and Astronomy, 116 Church St SE, MN 55455 Minneapolis, USA \\
$^{5}$ASTRON, Netherlands Institute for Radio Astronomy, Postbus 2, 7990 AA, Dwingeloo, The Netherlands \\
}

\date{Accepted 2018 December 10. Received 2018 December 10; in original form 2018 July 24}

\pubyear{2018}

\begin{document}
\label{firstpage}
\pagerange{\pageref{firstpage}--\pageref{lastpage}}
\maketitle

\begin{abstract}
% aims, methods, main results. Single paragraph < 250 words (< 200 words for Letters). No references.
The archetype FR-II galaxy Cygnus~A lies in a moderately rich cluster about to
undergo a major merger. We study the pre-merger Cygnus cluster environment using
smoothed particle hydrodynamics simulations constrained by 2Ms 
of \satellite{Chandra} observations of the hot intracluster medium. The 
observations constrain the total gravitating mass and concentration parameter, 
and the simulations provide the quiescent and merger-enhanced temperature profiles
of the pre- and post merger of the cluster excluding the central active galactic
nucleus. We present the first detailed model of the sub cluster north west of
Cygnus~A, named CygNW. We find a lower baryon fraction and higher concentration 
parameter for CygA than expected from known scaling relations in the literature. 
The model suggests the Cygnus cluster hosts a pre-merger with a progenitor mass
ratio of about 1.5:1 at the virial radius. We notice that the intra cluster medium
is heated as a result of the merger, but we find no evidence for a (pre-)merger shock in
the interstitial region between both cluster haloes. We attribute the merger-induced
heating to compression of the cluster outskirts. The smooth model obtained from our
simulations is subtracted from the observed cluster state and shows residual temperature 
structure that is neither hydrostatic nor merger-heated cluster gas. We speculate that
this residual heating may result from previous AGN activity over the last $\sim$~$100$~Myr.
\end{abstract}


%%%%%%%%%%%%%%%%%%%%%%%%%%%%%%%%%%%%%%%%%%%%%%%%%%

%%%%%%%%%%%%%%%%% BODY OF PAPER %%%%%%%%%%%%%%%%%%

\section{Introduction}
Cluster mergers, gas cooling, and galaxy feedback dominate the 
energetics of galaxy clusters in the current epoch. Mergers are expected
throughout cosmic time in the paradigm of hierarchical structure formation, while 
feedback becomes energetically important from redshift $z=2$ onwards
\citep[e.g.][]{2015ARA&A..53...51S}. Active Galactic Nuclei (AGN) provide one type
of such feedback. The Cygnus cluster prominently shows signatures of AGN feedback
as the brightest cluster galaxy (BCG) hosts the FR-II source Cygnus~A. 
In addition, the subcluster that this radio source is confined in is about to 
undergo a major merger \citep[][]{2013AN....334..346S}. Therefore, the Cygnus 
cluster provides a unique opportunity for an in-depth study of relative cluster 
energetics. Moreover, its proximity and extreme X-ray brightness combined with 
newly obtained, deep observations obtained at the excellent spatial resolution of
\satellite{Chandra} allows for a detailed study of the inner region directly
surrounding Cygnus~A as well as a study of the Cygnus cluster environment.

The well-known radio galaxy Cygnus~A has been studied extensively the last 
decades \citep[e.g.][]{1996cyga.book.....C, 1996A&ARv...7....1C}. The radio 
emission originates from accretion onto a supermassive black hole (SMBH) in the
BCG of the Cygnus cluster with an estimated black hole mass of
$(2.5 \pm 0.7) \times 10^{9}$~M\Sun \, \citep{2003MNRAS.342..861T}. Notable
X-ray features as a result of interaction of Cygnus~A with the surrounding
cluster gas are seen inside the cocoon shock in the inner $30-60$~kpc of the 
cluster. The central region shows two distinct X-ray jets with hot spots at the 
jet termini that can be used to estimate the AGN output energy. Other measurements
of the AGN energetics include computations of the work required to create the
observed X-ray cavities and, conversely, regions of higher density. The previous
deepest X-ray observation to date focussing on the large-scale structure of the 
Cygnus cluster is analysed by \citet{2002ApJ...565..195S}, and more recent work 
on the inner region can be found in \citet[][and companion papers]{2008MNRAS.388.1465S}, 
\citet{2010ApJ...714...37Y}, and \citet{2012A&A...545L...3C}.

The moderately rich Cygnus cluster in and of itself is also interesting and
well-studied. First reported as a powerful X-ray source in the \satellite{Uhuru}
catalog \citep{1972ApJ...178..281G}, the system has been observed by every major
mission since. An early rough mass estimate of $10^{14}$~M\Sun \, is given by 
\citet{1979ApJ...230L..67F} and confirmed by \citet{1984MNRAS.211..981A,
1987MNRAS.227..241A}. The authors further report that Cygnus~A lies in a cool 
core cluster (at the time called `a cluster with a strong cooling flow'), and
a $\beta$-model \citep{1978A&A....70..677C} fit to the intra cluster medium (ICM)
density profile yields a central density of ~$0.02$~cm$^{-3}$. Moreover, these
observations show a plume of hot intracluster gas extending to the north west, 
later confirmed by \citet{1994MNRAS.270..173C}, and now known to be a distinct 
subcluster. X-ray observations show that the cool-core cluster is undergoing a
major merger with the subcluster to the north west.

\begin{figure*}
    \centering
    \begin{minipage}{.49\textwidth}
        \includegraphics[width=\columnwidth]{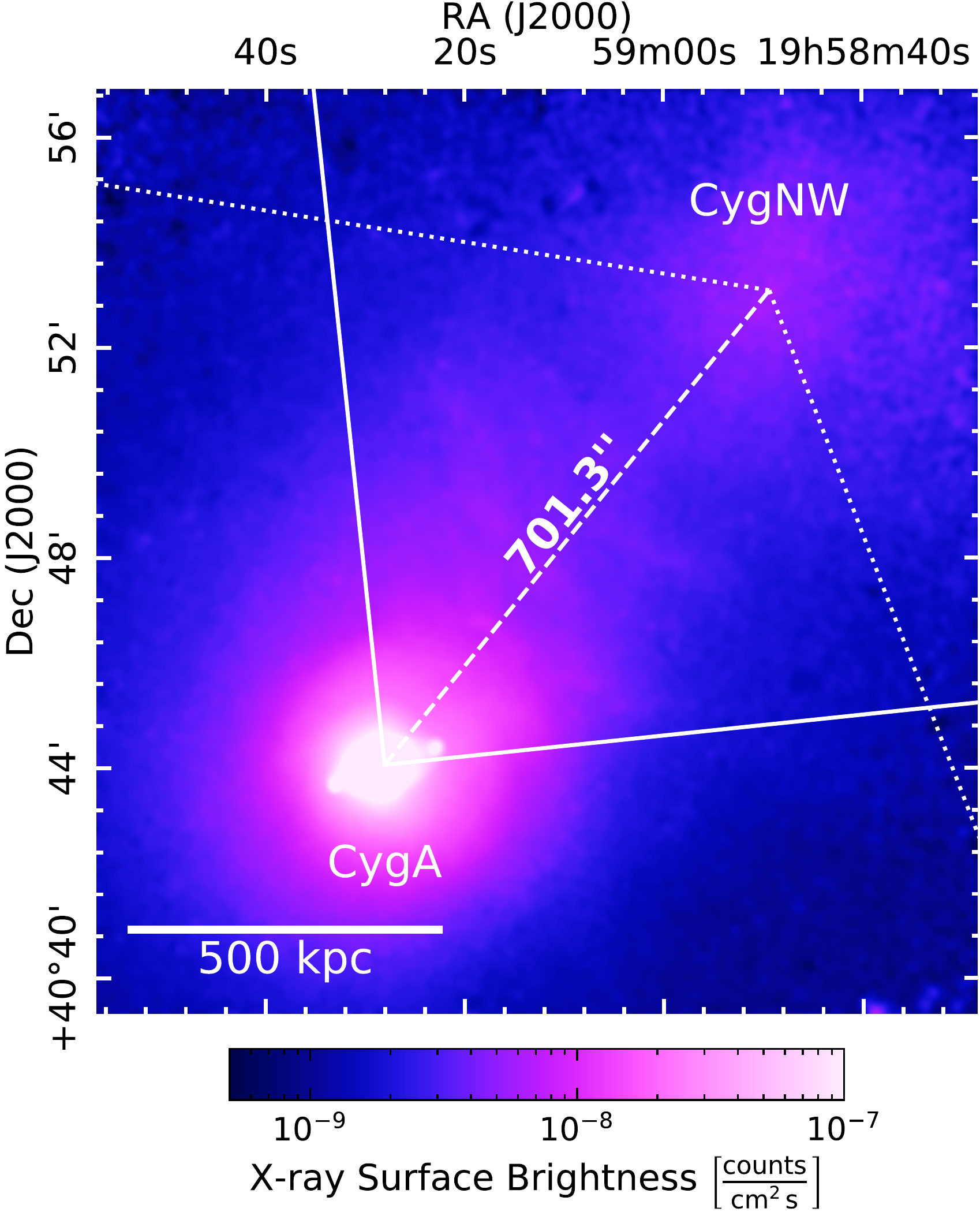}
    \end{minipage}
    \begin{minipage}{.49\textwidth}
        \includegraphics[width=\columnwidth]{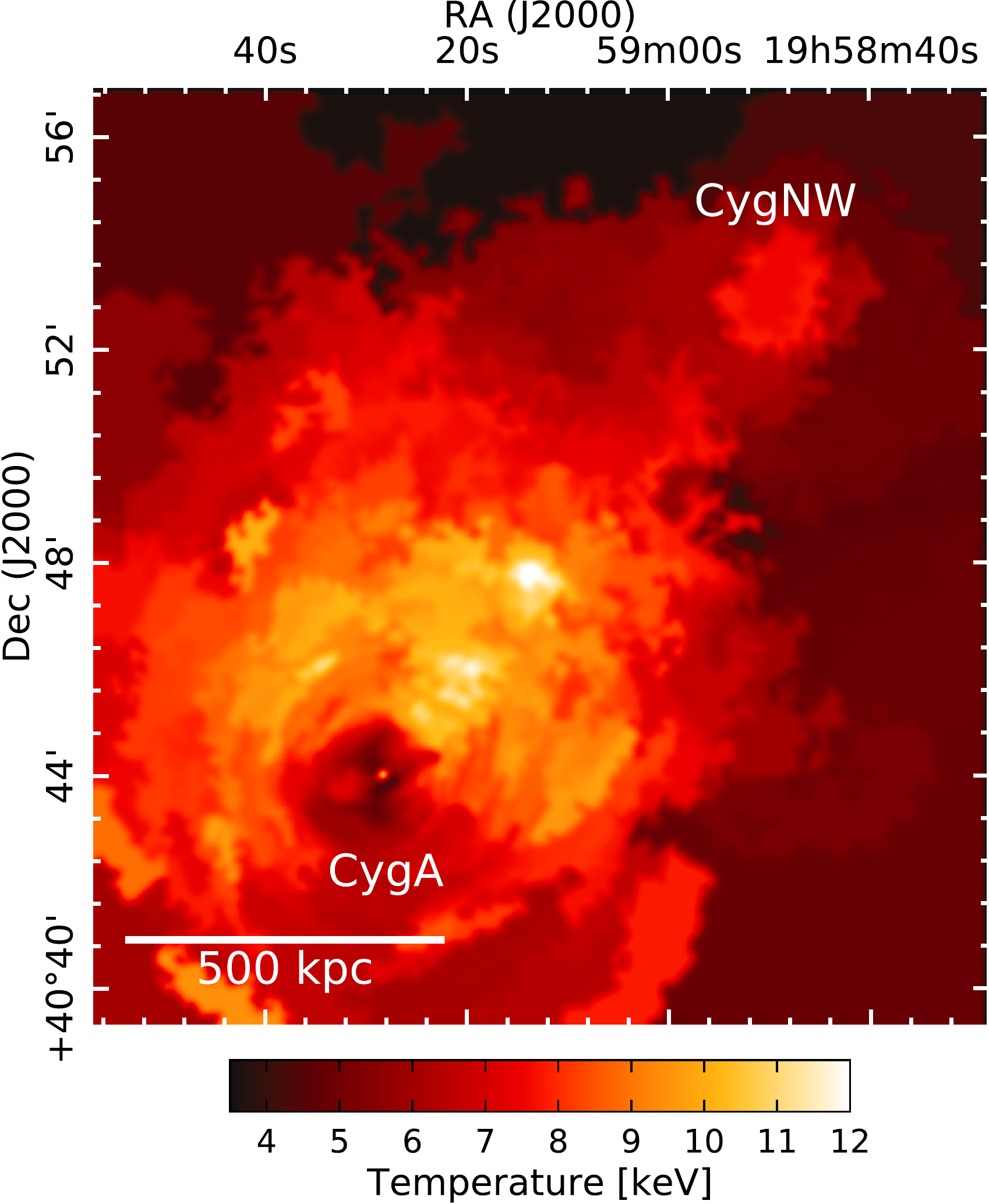}  
    \end{minipage}
    \caption{\emph{Left}: X-ray surface brightness (\satellite{Chandra}
             ACIS, $0.5-7.0$~keV, 2.2 Msec of exposure), convolved with a
             Gaussian smoothing kernel ($\sigma = 9$ pixels, $0.492''$/pixel). 
             We show a log stretch
             % with vmin=$5.0 \cdot 10^{-10}$ and vmax=$1.0 \cdot 10^{-7}$ counts/cm$^2$/s, 
             zoomed-in on the interstitial region. The
             dashed line shows a projection of the presumed merger axis to 
             a measured core separation of $701.3'' \sim (765 \, h_{70}^{-1}$~kpc).
             Excess emission is seen in-between the two haloes. The solid lines mark the
             boundaries of the wedges used to extract radial profiles
             for CygA, and the dotted line indicates these regions for CygNW.
             \emph{Right}: Observed temperature map in keV smoothed with a
             $9 \sigma$ Gaussian.}
    \label{fig:lss}
\end{figure*}

\citet{1999ApJ...521..526M} reported detection of a higher temperature in the 
region in-between both subclusters, indicative of a head-on pre-merger between 
(apart from the cool core in Cygnus~A) two nearly equal $4-5$~keV clusters that 
drives a merger shock into the ICM. \citet{2013AN....334..346S} further studied
the merger shock based on these data supplemented with \satellite{Suzaku}
observations. The latter shows an increase in redshift of Fe~K~line emission 
along the presumed merger axis\footnote{The dashed line in Figure~\ref{fig:lss}
(left) shows the projected merger-axis.}.
This increment is interpreted as a Doppler shift indicating that the cluster to
the north west is on the foreground with respect to- and falling in towards 
Cygnus~A. The measured radial component of the merger velocity is
$2650 \pm 900$~km/s, and the collision velocity is estimated from the temperature
jump as $2400-3000$~km/s. Combining both yields an inferred angle between the
merger-axis and the line-of sight of $54$\deg. 
% the redshift increases from $0.056 \pm 0.03$ to $0.070 \pm 0.05$ 

Dynamical modelling of optical observations shows an Abell richness of the cluster
of $2-4$, a velocity dispersion of $1581^{+286}_{-197}$~km/s, and indicates that 
the dynamical center of the spatial distribution is offset to the north west of Cygnus~A,
and the Cygnus~A galaxy has a redshift of $z=0.0562$ \citep{1997ApJ...488L..15O}.
The dynamical model supports the pre-merger state of the cluster estimated around 
$200-600$~Myr prior to core passage where the distance between both haloes is 
$460 h_{75}^{-1}$~kpc based on the galaxy surface density peaks, and the system
is seen at an inclination of $30-45$\deg~\citep{2005AJ....130...47L}. 
The authors further provide mass-estimates computed
using a projected mass estimator $M_{\textrm{PME}} = 4.4 \times 10^{15}$~M\Sun,
and from virial theorem $M_{\textrm{VT}} = 3.0 \times 10^{15}$~M\Sun. Prior 
literature indicates that the Cygnus cluster hosts a major merger with a mass
ratio of 2-3:1. To our knowledge, no weak-lensing studies or deeper optical 
observations of the cluster environment have been published to date.

One of the challenges with our new observations (Wise~et~al.~in~prep, see 
Section~\ref{sec:newchandra} below) is to differentiate between the contribution 
of the merger on the one hand, and on the other hand the effects of AGN output over the
past several hundred megayears in the same set of observations. To assist interpreting
the data we run idealised binary merger simulations of the ongoing merger in the
Cygnus cluster where we exclude the AGN and its effect on the ICM. The simulations can 
constrain the quiescent hydrostatic temperature of the ICM as well as merger-heated
temperature profiles. The goal is to understand the Cygnus cluster environment,
particularly the properties of both merging subclusters as well as the energetics of
the merger.

Newly obtained, detailed radial profiles of both subclusters constrain the properties
of the merger in the Cygnus cluster. We adopt the relevant \satellite{Chandra}
observations from Wise~et~al.~(in~prep) that are obtained as part
of a recent multi-wavelength observational campaign of persistent AGN activity in the
merging Cygnus cluster in section~\ref{sec:newchandra}. The X-ray observations as well as
the optical dynamical model suggest a head-on merger between two nearly equal clusters 
of galaxies. This is also seen in the newly obtained deep \satellite{Chandra} observations. 
Our cluster model and the inferred halo parameters are presented in section~\ref{sec:model}, 
followed by the numerically sampled clusters in section~\ref{sec:ics}. We present simulations 
of the head-on merger scenario in section~\ref{sec:merger} and discuss the interpretation of
the simulations in section~\ref{sec:discussion}. We conclude in section~\ref{sec:conclusions}.

Throughout this paper we assume a concordance $\Lambda$CDM cosmology with $H_{0} = 70$~km/s/Mpc,
$\Omega_{\Lambda} = 0.7$, and $\Omega_{M} = 0.3$. This yields an angular diameter
of $1''\sim 1.091$~kpc and a critical density of $\rho_{\text{crit}} = 9.7
\times 10^{-30}$ g/cm$^{3}$ at the redshift of the Cygnus cluster.

\section{\satellite{Chandra} Constraints on the Cygnus Merger Simulations}
\label{sec:newchandra}
New \satellite{Chandra} ACIS observations of the Cygnus cluster were obtained
as part of a recent multi-wavelength observational campaign 
\citep{2014cxo..prop.4448W}. An overview of the data analysis of the two 
mega seconds of \satellite{Chandra} data is outlined by
Wise~et~al.~(in~prep). A detailed discussion of
the non-thermal emission originating from the central region surrounding the AGN
is published by \citet[][]{2018MNRAS.478.4010D}. \citet{2016MNRAS.463.3143M}
analysed LOFAR data on the emission originating from the hotspots at the 
jet termini, and \citet{2016ATel.9495....1P} obtained new JVLA observations of
Cygnus~A. 

\subsection{X-ray map of the Cygnus cluster}
The left panel of Figure~\ref{fig:lss} shows the \satellite{Chandra} X-ray surface 
brightness ($0.5-7.0$~keV) with a total exposure time of 2.2~Msec. The Cygnus cluster
hosts two distinct sub clusters. We refer to as the sub cluster that the radio source 
Cygnus~A is confined as `CygA', and following Wise~et~al.~(in~prep), 
the distinct subcluster to the north west is named `CygNW'. CygA appears roughly spherical 
although departure from perfect spherical symmetry in the direction of the western jet can 
be seen. CygNW appears slightly elongated towards - and is much fainter
than - CygA. In fact, CygNW is only visible after significant smoothing with a 
$\sigma = 9$ pixels ($0.492''$/pixel) Gaussian smoothing kernel.

We focus on the interstitial region in particular where we observe filamentary 
structure. This excess surface brightness in-between CygA and CygNW is clearly
seen along the presumed merger-axis (indicated by the dashed line). In this work 
we investigate whether the excess could be merger-induced or signatures of previous 
AGN activity. The measured projected core separation, the distance between both
halo centroids, is $701.3'' \sim (765 \, h_{70}^{-1}$~kpc). 

\subsection{Temperature structure}
The temperature structure of the Cygnus cluster is presented in Figure~\ref{fig:lss}
(right). Wise~et~al.~(in~prep) presents a detailed description of 
the reduction and analysis of the observations; here we summarise the key aspects. All 
observations are reprocessed with \code{ciao}~4.9 and \code{caldb}~4.7.4
\citep{2006SPIE.6270E..1VF}. The observed temperature map is created using the
\code{Contbin} algorithm \citep{2006MNRAS.371..829S} to generate two-dimensional bins.
\code{Contbin} requires that each bin has a surface brightness with a constant signal-to-noise 
ratio of $75$ and increases the bin until this condition is met. 

For each bin \textit{specextract} is used to extract spectra, and the \code{Sherpa}
\citep{2001SPIE.4477...76F} fitting package within \code{ciao} is then used to spectral 
fit a single-temperature \code{apec} model \citep{2001ApJ...556L..91S} multiplied with
a \code{tbabs} absorption model \citep{2000ApJ...542..914W}. A column density of 
$N_H = 3.1 \times 10^{21}$~cm$^{-2}$ for absorption due to neutral hydrogen in the
Galactic plane is obtained by averaging values adopted from the Leiden/Argentine/Bonn (LAB)
\citep{2005A&A...440..775K} and \citet{1990ARA&A..28..215D} surveys. Finally, we convolve the 
resulting temperature map with a Gaussian kernel ($\sigma = 9$) to smooth over AGN interactions
in the central regions while simultaneously enhancing the cluster-scale features. The cool
core of CygA is visible in the temperature map, while the central region of CygNW is hot
compared to its direct surroundings. 

\subsection{Beta model fit to \satellite{Chandra} radial profiles}

\begin{figure*}
    \includegraphics[width=.47\textwidth]{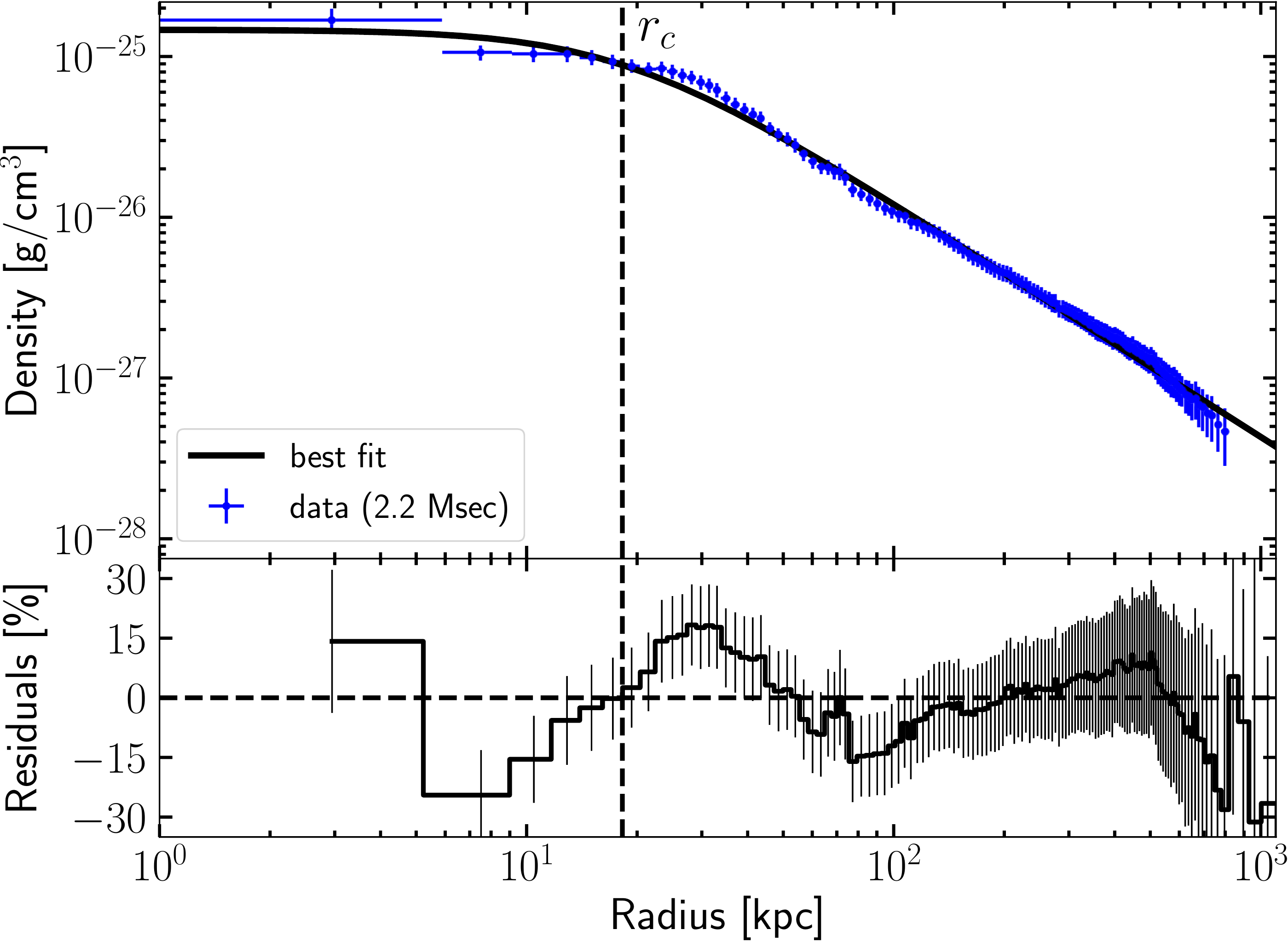}
    \includegraphics[width=.47\textwidth]{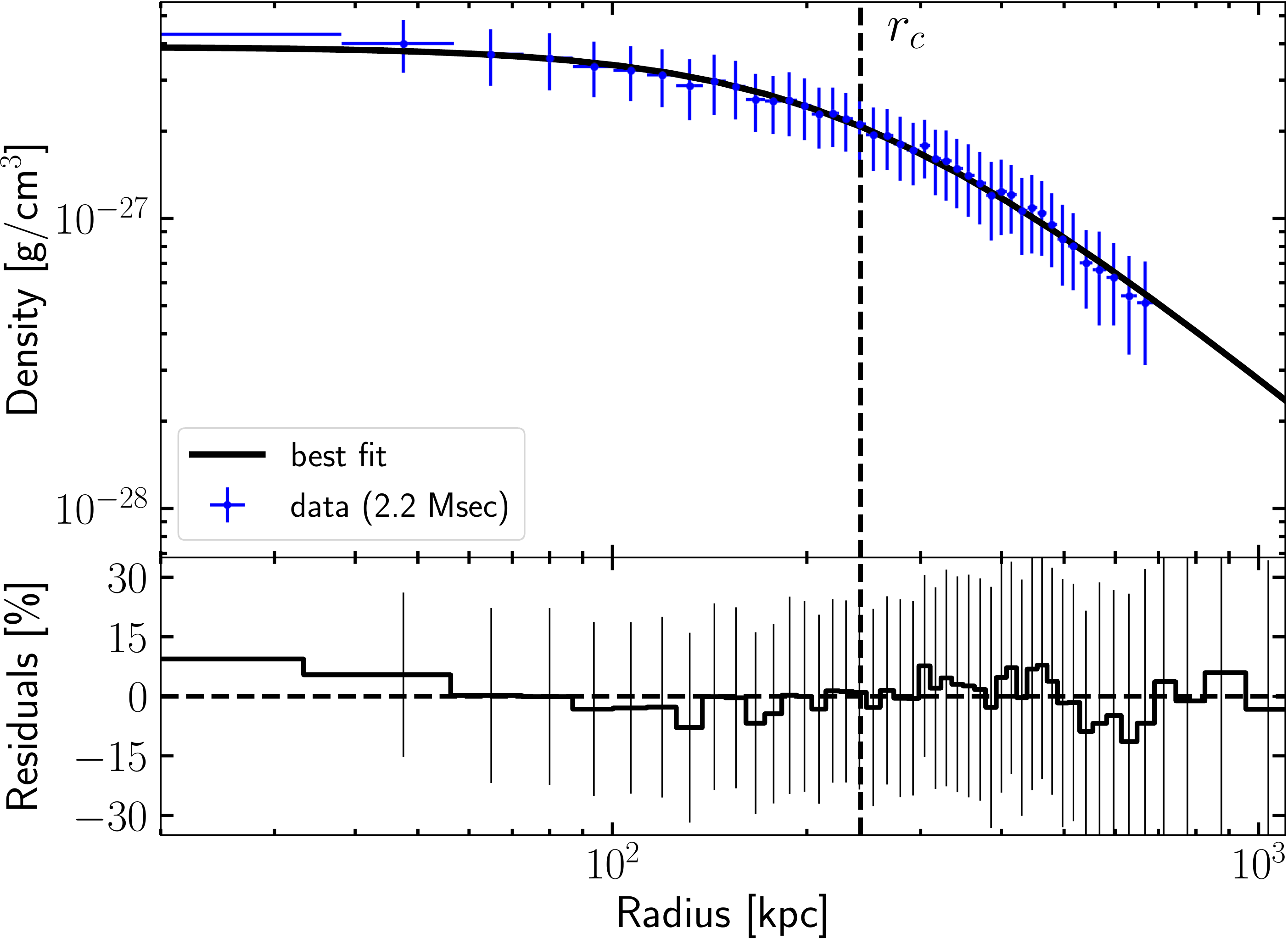}
    \caption{$\beta$-model fit to the observed, adaptively-binned X-ray surface
             brightness wedge excluding the merger-axis, centered on CygA (left)
             and CygNW (right).}
    \label{fig:betamodel_fit}
\end{figure*}

Radial profiles are extracted by Wise~et~al.~(in~prep);
here we summarise the relevant steps in the data reduction only. The interstitial
region of enhanced surface brightness is cut out in efforts to create `quiescent'
radial profiles of CygA and CygNW. This is done to minimise contamination
of whichever physical phenomenon causes the excess emission. The solid lines
centered on CygA mark the two regions used to extract the radial profiles. The wedge
enclosing the merger-axis is coined the `merger-region' and spans $6-96$\deg~measured
from west to the north, while its inverse is dubbed the `quiescent' or average region. 
The dotted lines around the merger-axis indicates the region cut out to obtain the
quiescent radial profile of CygNW.

Wise~et~al.~(in~prep) subdivide each wedge in partial annuli, bin 
them in such a way that each annulus contains the same number of counts. Finally, the
authors fit an \code{apec} model to the extracted spectra to obtain the density and 
temperature values. The adaptive binning algorithm requires a signal-to-noise ratio 
of 273 (75) for CygA (CygNW), thus the radial profiles have uniform error bar sizes. 
We present our $\beta$-model \citep{1978A&A....70..677C} fit to the radial profile within the
`quiescent' wedge in Figure~\ref{fig:betamodel_fit} for CygA (left panel) and for
CygNW (right panel). The best fit $\beta$-model parameters are presented in 
Table~\ref{tab:ics} (top).

% 2.2Msec dataset (20180313). For betamodel fit cut and uncut give same results.
\begin{table}
    \centering
    \caption{Best fit parameters used as initial conditions to set up numerical
             representations of the CygA and CygNW haloes. The $\beta$-model 
             parameters are constrained by the \textit{Chandra} density profile, 
             while the remaining quantities are inferred by fitting the total
             hydrostatic temperature (assuming dark matter follows the NFW-profile)
             to the observed temperature profile. We use the model with an additional 
             cut-off in the NFW-profile and $\beta$-model. The value
             $c_{\text{NFW}} (M_{200})$ is computed using the scaling 
             relation of \citet{2008MNRAS.390L..64D}.
         }
    \label{tab:ics}
    \begin{tabular}{llcccc} 
        \hline
        Parameter & & \multicolumn{2}{c}{CygA} & \multicolumn{2}{c}{CygNW} \\
        \hline
        $\rho_0$ & ($10^{-26}$ g/cm$^{3}$) & \multicolumn{2}{c}{$14.68 \pm 0.61$}    & \multicolumn{2}{c}{$0.39 \pm 0.01$} \\   %2Msec
        $r_c$    & (kpc)                   & \multicolumn{2}{c}{$18.22 \pm 0.78$}   & \multicolumn{2}{c}{$242.1 \pm 9.86$} \\    %2Msec
        $\beta$  &                         & \multicolumn{2}{c}{$0.486 \pm 0.004$} & \multicolumn{2}{c}{$0.608 \pm 0.02$} \\     %2Mse®
        $\chi_{\text{min}}^2/$dof &        & \multicolumn{2}{c}{$63.6 \, / \, 127$} & \multicolumn{2}{c}{$1.4 \, / \, 39$} \\  %2Msec
        \hline
        % &                                               & uncut & cut      & uncut & cut      \\
        $\rho_{\text{0dm}}$ & ($10^{-25}$ g / cm$^{3}$) &       & $1.82$   &       & $0.23$  \\
        $c_{\text{NFW}}$ &                              &       & $7.18$   &       & $2.87$  \\
        $c_{\text{NFW}} (M_{200})$ &                    &       & $3.35$   &       & $3.48$  \\
        $bf_{200}$ & (per cent)                         &       & $7.01$   &       & $7.35$  \\
        $M_{200}$ & ($10^{14}$ M\Sun)                   &       & $7.36$   &       & $4.99$  \\
        $M_{200gas}$ & ($10^{14}$ M\Sun)                &       & $0.52$   &       & $0.37$  \\
        $M_{500}$ & ($10^{14}$ M\Sun)                   &       & $5.66$   &       & $3.16$  \\
        $r_{200}$ & (kpc)                               &       & $1831$   &       & $1609$  \\
        $r_{500}$ & (kpc)                               &       & $1235$   &       & $1018$  \\
        $r_{s}$ & (kpc)                                 &       & $255.0$  &       & $560.9$ \\
        $r_{\text{sample}}$ & (kpc)                     &       & $6865$   &       & $2413$  \\
        $r_{\text{cut}}$ & (kpc)                        &       & $854.7$  &       & $987.3$ \\
        \hline
    \end{tabular}
\end{table}

\section{Cluster Model}
\label{sec:model}
We adopt the cluster model of \citet[hereafter D14]{2014MNRAS.438.1971D} and 
\citet[D17]{2017MNRAS.471.4587D}. D17 use the \citet*[][NFW]{1995MNRAS.275..720N} 
profile to describe the dark matter density as a function of radius. An additional
cut-off is used because the profile does not converge for $r \rightarrow \infty$.  
The integral with the cut-off density profile has to be solved numerically.
The expression for the cutoff NFW density is
\begin{align}
    \rho_{\text{dm}}(r) &= \rho_{\text{0dm}} \left[ \frac{r}{r_s} \left(1+\frac{r}{r_s}\right)^2 \right]^{-1}
          \left[ 1 + \left(\frac{r}{r_{\text{sample}}}\right)^3 \right]^{-1} \quad , \label{eq:nfw}
\end{align}
where $r_s$ is the scaling radius, and $\rho_{\text{0dm}}$ is a scaling parameter
for the density profile. As per D14, the sample radius $r_{\text{sample}}$ is
set to half the size of the simulation box, $L_{\text{box}} = 3.75 \, r_{200}$
for the primary halo (CygA), and to $1.5 \, r_{200}$ for the additional halo 
(CygNW). The virial radius $r_{200}$ is defined as the radius at which the average
(spherical) density equals two hundred times the critical density of the Universe
at the redshift of the cluster. The concentration parameter of the dark matter 
halo is defined as the ratio of the virial radius to the scaling radius, 
$c_{\text{NFW}} = r_{200} / r_s$, and is a free parameter in our fit.

The baryonic content of clusters consists of the intra cluster medium and 
the stellar component. We neglect the latter, and for the ICM we define the baryon
fraction as $b_f = M_{\text{gas}} / (M_{\text{gas}} \, + \, M_{\text{dm}})$ and 
the gas density profile follows the $\beta$-model \citep{1978A&A....70..677C}. D17 adopt
an additional cut-off in the $\beta$-model to ensure that the local baryon fraction
in the numerical implementation stays below unity throughout the simulation box
for stability, with 

\begin{align}
    \rho_{\text{gas}}(r) &= \rho_{0\text{ICM}} \left[ 1 + \left(\frac{r}{r_c}\right)^2 \right]^{-3 \beta /2}
    \cdot \left[ 1 + \left(\frac{r}{r_{\text{cut}}}\right)^3 \right]^{-1} \quad , \label{eq:betamodel}
\end{align}

\noindent with core radius $r_c$ and power-law slope $\beta$ outside the core. The cut-off
radius is parametrized as a fraction of the virial radius, where D17 finds 
$r_{\text{cut}} / r_{200} \sim 1.7$ in the Perseus cluster.

Finally, D14 and D17 assume that both progenitors are spherically symmetric and in hydrostatic 
equilibrium where self-gravity is balanced by gas pressure.

\subsection{Progenitor properties}
\label{sec:progenitor}
We adopt the view that the Cygnus cluster is in a pre-merger state roughly $200-600$~Myr
prior to core passage \citep{2005AJ....130...47L}. The undisturbed X-ray morphology 
and the prominent cool core of CygA further supports the pre-merger conclusion 
\citep{2005AJ....130...47L}. We thus assume that the progenitor haloes are
described by the current observations, specifically by the radial profiles within 
the quiescent wedge.

In addition to the radial density profiles already presented in Figure~\ref{fig:betamodel_fit}, 
Wise~et~al.~(in~prep) has also extracted radial temperature
profiles. We first compute the total gravitating mass from the observed density
profile by assuming a fixed baryon fraction at the virial radius. The temperature
profile then follows from hydrostatic equilibrium, and the pressure can be computed
by adopting the ideal gas law as equation of state. We $\chi^2$ fit the computed
hydrostatic temperature profile to the observed temperature profile where the 
free parameters are $r_{200}$, $c_{\text{NFW}}$, and $b_f$. Decreasing the baryon
fraction results in a higher overall temperature of the cluster, while the
concentration parameter determines the amount of dark matter within the scaling
radius. The latter effectively shifts the peak of the temperature profile
inward or outward by respectively increasing or decreasing the concentration parameter.
For CygA we only take into account $50$~kpc$ \, < r < 500$~kpc. The cocoon shock
around Cygnus~A is observed in the inner $\sim 30-60$~kpc (depending on the direction),
thus, the temperature structure in this region is highly influenced by the AGN rather
than only showing the hydrostatic temperature of the cluster.

Based on these observations we find $\beta$-model properties for CygA that are consistent 
with earlier findings of \citet{2001ApJ...556L..91S}, who reported 
$r_c \approx 18$'' ($19.3$~kpc), $\beta \approx 0.51$ for the cluster gas in the inner
$300$''. We find a core radius of $18.23 \pm 0.78$~kpc, $\beta = 0.486 \pm 0.004$ 
within $1$~Mpc. Furthermore, in order to match the hydrostatic temperature to the
observed temperature profile we find a low baryon fraction and high concentration 
parameter of $bf_{200} = 7.0$ \% and $c_{\text{NFW}} = 7.18$. Using the
\citet{2008MNRAS.390L..64D} scaling relation between the concentration parameter
and total mass, we expect a value of value $c_{\text{NFW}} (M_{200}) \sim 3.35$.
Although the scaling relation is not tightly constrained, the concentration that 
we find for CygA falls in the upper part of the scatter in the scaling relation.
A baryon fraction of $0.039 - 0.055$ has previously been reported by 
\citet{2001ApJ...556L..91S}, a bit lower than our findings.

The progenitor properties are presented in Section~\ref{sec:ics}.
Table~\ref{tab:ics} shows the inferred parameters with the additional cut-off 
in the density profiles. Our mass estimate is given at two characteristic radii, 
both at the virial radius $r_{200}$ and at $r_{500}$ to assist comparing the 
Cygnus cluster haloes with known scaling relations in the literature. The 
corresponding radial profiles are plotted in Figure~\ref{fig:donnertsubplot_cygA} 
for CygA, respectively Figure~\ref{fig:donnertsubplot_cygNW} for CygNW. The 
observed data points are shown in blue. The upper and lower right panel show the 
hydrostatic mass and pressure, both of which are computed from the observed 
density and temperature profiles. The black lines show the analytical profiles 
where we note that the $\beta$-model and hydrostatic temperature are fitted to 
the observations. The mass- and pressure profiles self-consistently lie on top
of the respective observed profiles. Finally, the green points show the numerically 
sampled profiles which we further address in section~\ref{sec:ics}.

\subsection{Mass ratio}
\label{sec:massratio}
\begin{figure}
    \includegraphics[width=\columnwidth]{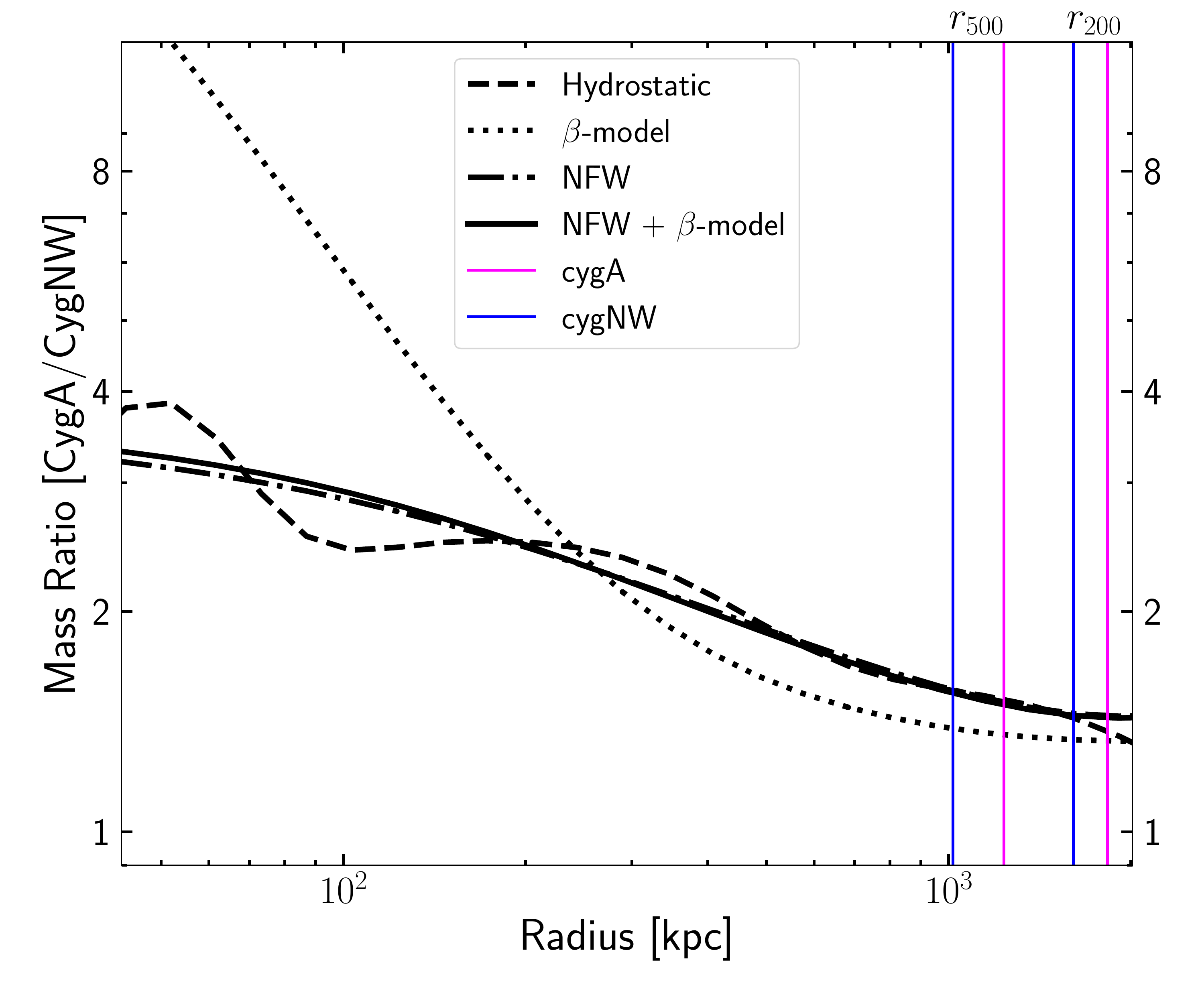}
    \caption{Mass ratio as computed from the observations (hydrostatic mass
             equation, dashed line) versus the mass ratio found by modelling 
             the cluster environment. The solid line shows the ratio of total
             mass while the baryonic and dark mass ratio is shown by the botted,
             respectively, dash-dotted line. The vertical lines show $r_{500}$ 
             and $r_{200}$ of CygA (magenta), and CygNW (blue). Note that the 
             virial radii lie well outside the observed data range.}
    \label{fig:massratio}
\end{figure}

The dynamical model of \citet{2005AJ....130...47L} shows that the mass ratio 
in the Cygnus cluster is at most of order 2:1, where CygA is the more massive
cluster. Our modeling, on the other hand, suggests that the progenitors have a 
mass ratio of \textbf{1.5:1} at the virial radius $r_{200}$ as seen in
Table~\ref{tab:ics}. We show the mass ratio as a function of radius in 
Figure~\ref{fig:massratio}.

Following Wise~et~al.~(in~prep), we adopt the hydrostatic mass
profile of \citet[][eq. 6]{1980ApJ...241..552F} to compute the observed total 
mass as a function of radius given the observed density and temperature profiles
and their derivatives. We obtain the latter by smoothing the observed profiles,
fitting a cubic spline and taking the derivative of the spline. We show the 
resulting mass ratio with a dashed line and note that the adopted equation 
only assumes hydrostatic equilibrium and spherical symmetry, but requires no
assumptions for the shape of the dark matter halo. The dotted line 
shows the mass ratio of the baryons in the ICM, and the mass ratio of the dark
component is shown by the dash-dotted line where we assume that the dark matter
halo is described by the NFW-profile. We note that the inferred virial radii
fall well outside the range of observed radii, and the sum of the virial radii
of CygA and CygNW is significantly larger than the observed core separation, 
further discussed in section~\ref{sec:discussion}.

\section{Initial Conditions}
\label{sec:ics}
The cluster model is implemented in the \code{OpenMP} parallel code 
\code{Toycluster}\footnote{\href{https://github.com/jdonnert/Toycluster}
{https://github.com/jdonnert/Toycluster}}, first presented by D14 while the latest
changes are published in D17. Two minor modifications with respect to the latter
have been made for our purposes to sample the initial conditions for CygA and CygNW 
as given in Table~\ref{tab:ics}. We allow for an individual baryon fraction for both 
haloes, and for different cut-off radii per halo to have our fiducial cluster match 
the observations as accurately as possible.

\begin{figure*}
\includegraphics[width=\textwidth]{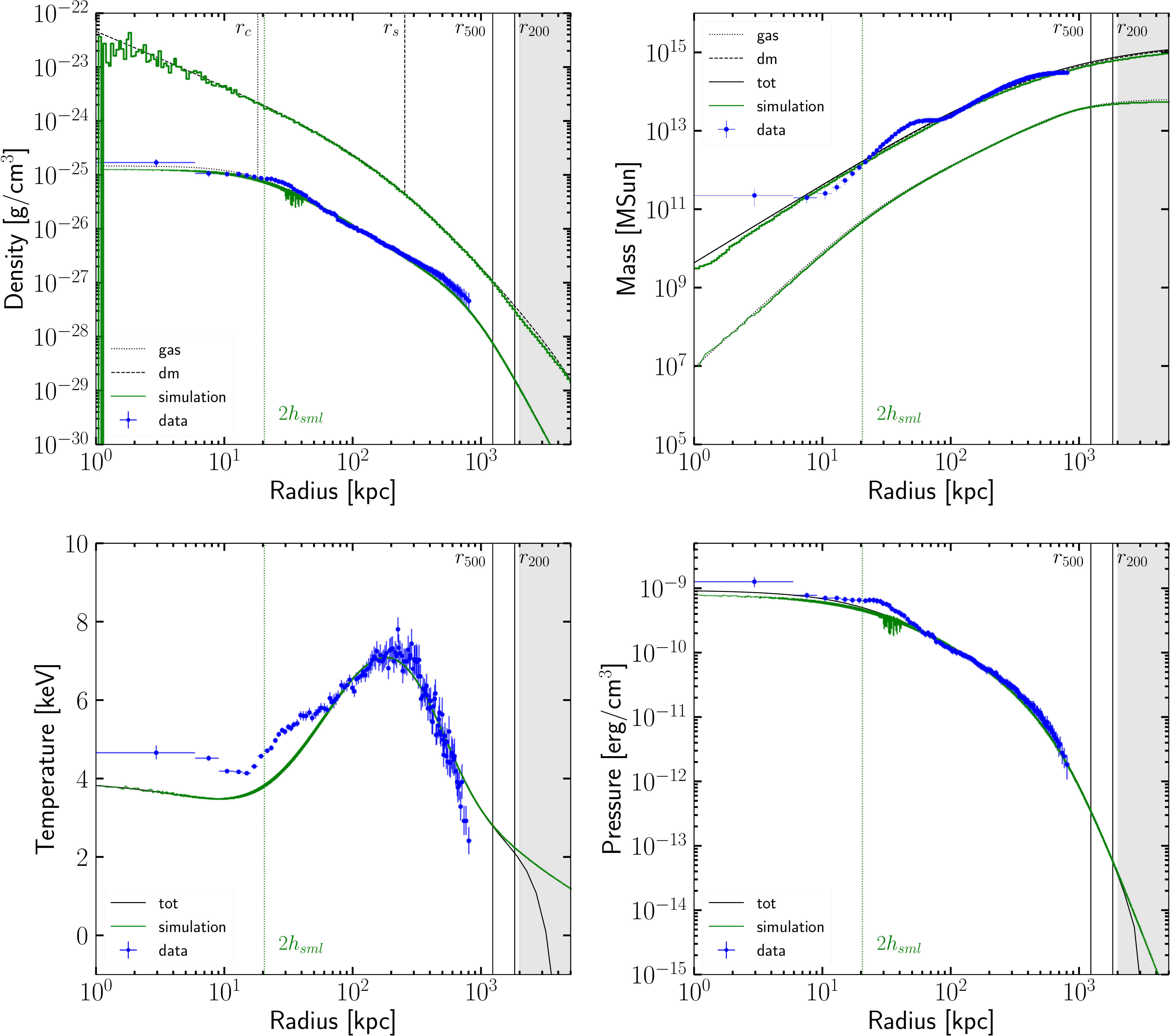}
    \caption{Comparison between fiducial radial profiles of CygA of the initial 
             conditions (shown in green) and the observed `quiescent' radial profiles (blue).
             We show the density profile (upper left), mass profile (upper right), 
             temperature profile (lower left), and pressure profile (lower right).
             The $\beta$-model is fitted to the observed density profile, and the
             dark matter (NFW) profile is inferred under the assumption of a
             fixed baryon fraction at the virial radius $r_{200}$. The core radius
             of the $\beta$-model and the scaling radius of the NFW profile are indicated
             by the vertical dotted, respectively, dashed line, and the vertical
             green line indicates the resolution scale.}
    \label{fig:donnertsubplot_cygA}
\end{figure*}

\begin{figure*}
\includegraphics[width=\textwidth]{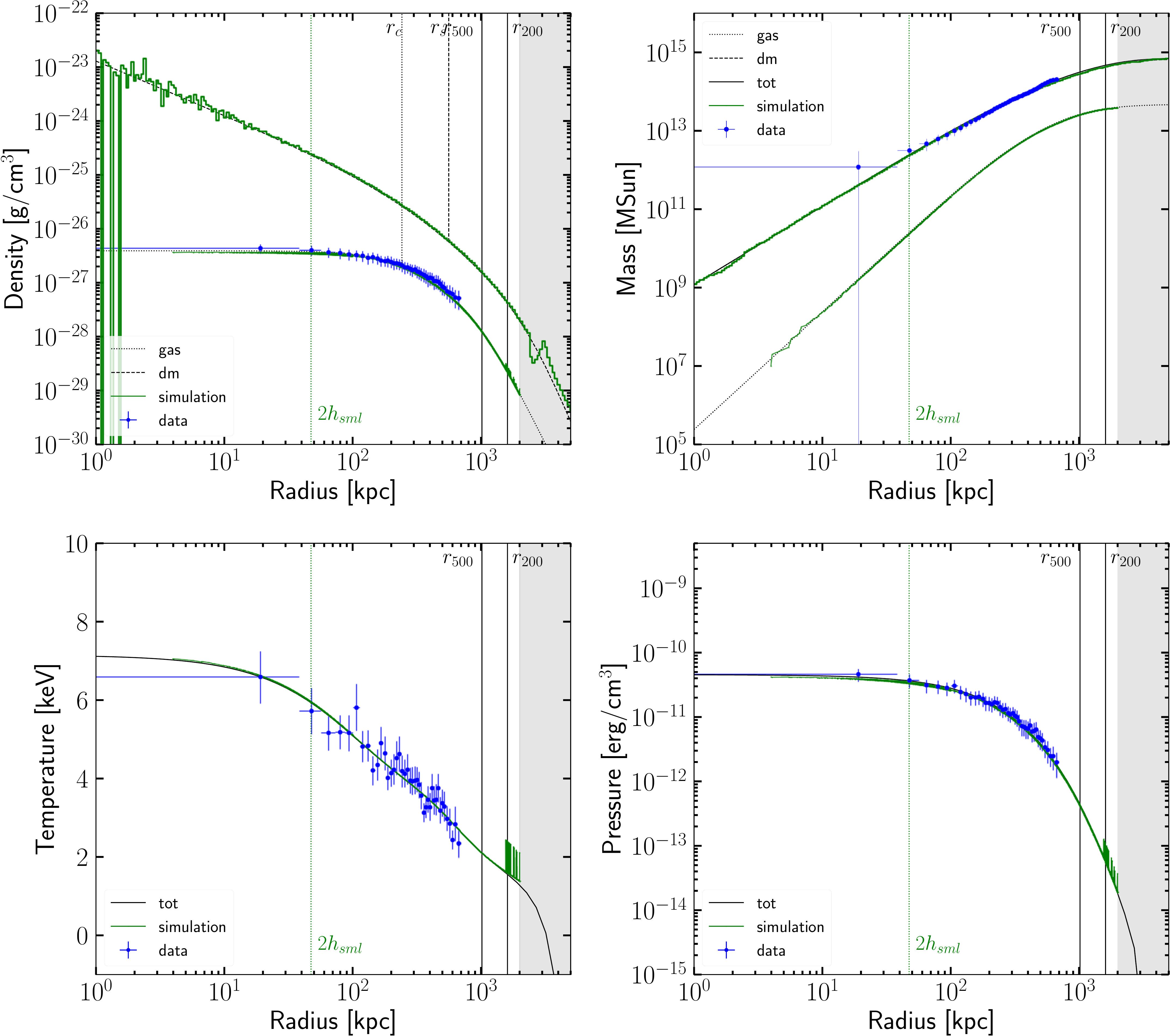} 
    \caption{Comparison between fiducial radial profiles of CygNW of the initial 
             conditions (shown in green) and the observed `quiescent' radial profiles (blue).
             We show the density profile (upper left), mass profile (upper right), 
             temperature profile (lower left), and pressure profile (lower right).
             The $\beta$-model is fitted to the observed density profile, and the
             dark matter (NFW) profile is inferred under the assumption of a
             fixed baryon fraction at the virial radius $r_{200}$. The core radius
             of the $\beta$-model and the scaling radius of the NFW profile are indicated
             by the vertical dotted, respectively, dashed line, and the vertical
             green line indicates the resolution scale.}
    \label{fig:donnertsubplot_cygNW}
\end{figure*}

The initial Poisson sampling error of order $21\%$ is reduced to $<3\%$ as the result
a relaxation loop based on the weighted voronoi tesselation (WVT) algorithm of 
\citet[][]{2012arXiv1211.0525D}. In each relaxation step a fiducial force 
is computed to push particles apart to improve the particle sampling and reduce
the mean sampling error, see D17 and Arth~et~al.~(in~prep) for details. 
Individual clusters are then stable under hydrostatic equilibrium for more than 5~Gyr.
Moreover, the numerically sampled profiles follow the analytical requirements without
much scatter as a result of the WVT relaxation. The numerically sampled radial
profiles are shown as the green points and lines in Figure~\ref{fig:donnertsubplot_cygA}
for CygA, respectively, Figure~\ref{fig:donnertsubplot_cygNW} for CygNW. The mean 
smoothing length of the particles within the inner $100$~kpc is shown as a vertical
dotted green line to indicate the kernel resolution scale. The shaded grey area
indicates the boundary region where the model is no longer valid. 

\subsection{Progenitor orbit}
\label{sec:orbit}
We sample the cluster with desired parameters in a simulation box and place 
the haloes on a collision course assuming that the merger trajectory is described 
by a parabolic orbit. 

Following D17, the initial velocity is parametrized using the zero-energy orbit
fraction $X_E$. This parameter lies between zero and one. The value $X_E=0$ 
corresponds to the case where both clusters are initially at rest at infinite distance.
Conversely, for $X_E=1$ all available potential energy is converted to 
kinetic energy. The first relevant snapshot of the simulations is when the merger 
starts. Therefore both haloes are placed in the box at a distance such that their 
virial radii touch. No hard constraints on the impact parameter are given in the 
literature, but the merger scenario of \citet{1999ApJ...521..526M} suggests a 
head-on collision. For this reason the impact parameter of the Cygnus merger 
simulations is set to zero (head-on). Although we are not bound to this assumption,
we did not further investigate this parameter due to the lack of observational constraints.
Three initial merger velocities are chosen with values of $X_E \in
[0.25, 0.50, 0.75]$, thus, three different initial condition files are generated.
This corresponds to an approach velocity $v_{\text{initial}} \in [650, 1301, 1952]$~km/h.
For all simulations we also investigate the inclination $i$ of the system. A
value of $i=0\deg$~corresponds to the merger occurring in the plane of the sky,
while $i=90\deg$~indicates a merger along the line of sight. The inclination 
angle can be chosen after the simulation run to create two-dimensional projections
of physical quantities as part of the post-processing. We chose to investigate
the temperature structure of the simulated cluster for inclination angles of
$i \in [0, 15, 30, 45, 60, 75]$. The resulting merger simulations are presented
in the following section.

\section{Results}
\label{sec:merger}
We use the \code{TreeSPH} code \code{Gadget3} \citep{2005MNRAS.364.1105S,
2009MNRAS.398.1678D, 2016MNRAS.455.2110B} with the bias-corrected Wendland~C6
kernel with $295 \pm 0.01$ neighbours \citep{2012MNRAS.425.1068D}. The simulation
runs are performed on the Dutch Compute Cluster Lisa and on the Freya compute 
cluster at the Max Planck Computing and Data Facility (MPCDF), using $20$ million 
particles split up equally between the dark matter and the gas particles. This 
results in an effective mass resolution of $9.84 \times 10^6$ M\Sun \, for SPH, 
and $1.77 \times 10^8$ M\Sun \, for DM particles. Snapshots are written
every $10$ Myr for a simulation run of $3$~Gyr without comoving integration as
the redshift of the Cygnus cluster is $0.0562$, and the gravitational softening 
is set to one seventh of the mean particle separation, or $3.61$~kpc.

\subsubsection*{Temperature increase and puff-up}
We confirm that both clusters are stable over the course of the simulation by 
plotting the radial profiles of both haloes at every time step. We notice that 
the temperature profile in the inner $\sim300$~kpc of fiducial CygA rapidly 
increases by $\sim10\%$ during the first $\sim 230$~Myr of the simulation run.
After this initial rapid increase the temperature profile stabilises. The 
temperature of CygNW ($r<50$~kpc) also increases slightly, but remains within 
the \satellite{Chandra} errorbars. The increase in central temperature is not 
observed when both fiducial subclusters are sampled individually (i.e. in their 
own simulation box).

In addition, the scatter of the temperature profile increases over the course of the
simulation run although the initial conditions started with radial profiles that 
closely follow the analytical profiles without much scatter as a result of the WVT 
relaxation step. The density, on the other hand, remains well-constraint with a small
scatter.
%20170115T0906

\begin{figure*}
    \includegraphics[width=.47\textwidth]{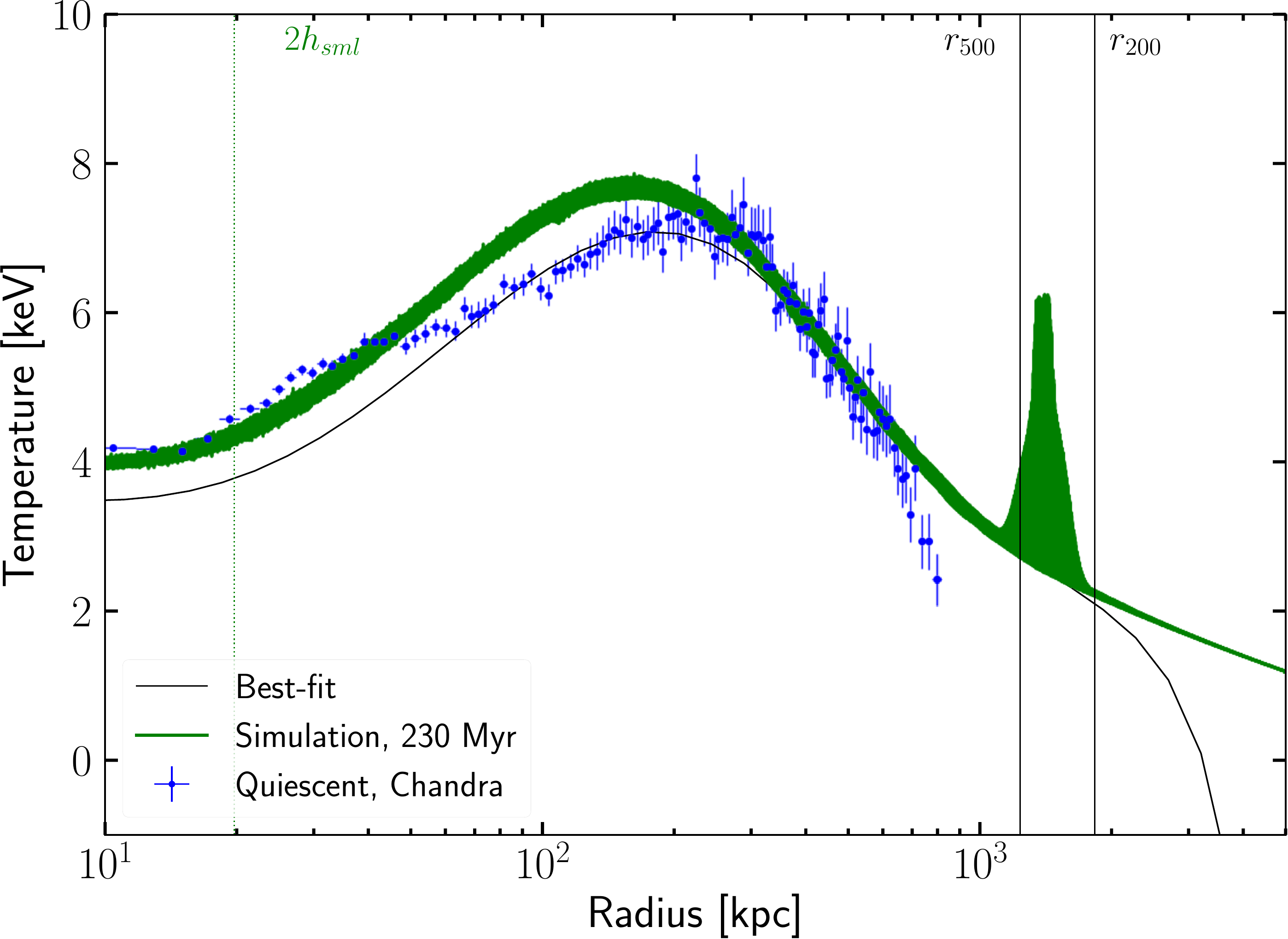}
    \includegraphics[width=.47\textwidth]{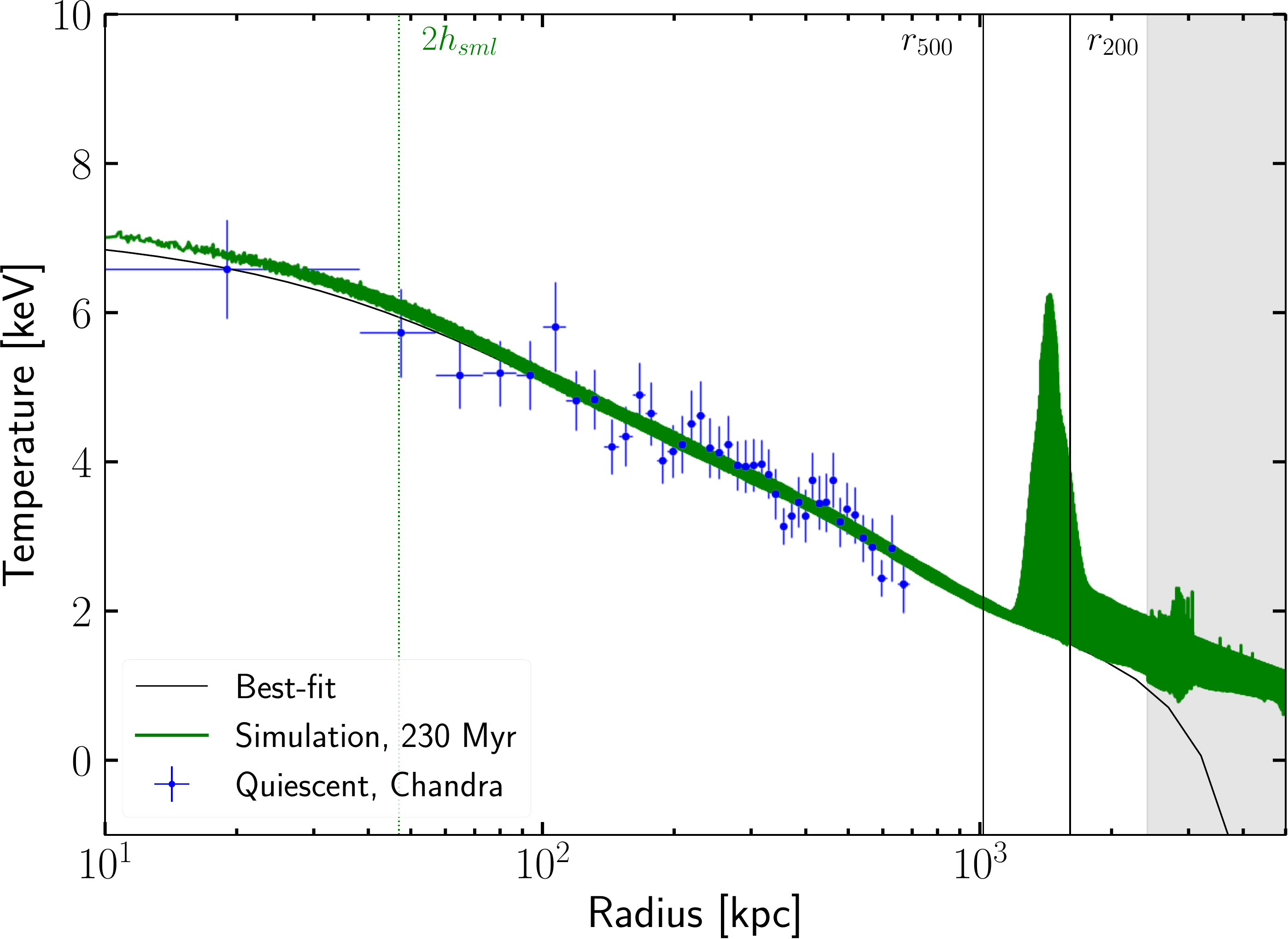}
    \caption{Temperature structure of the $X_E = 0.50$ simulation after
             $230$~Myr. \emph{Left:} fiducial CygA overplotted on the quiescent
             temperature profile of CygA. For clarity we also show the best-fit
             model (dotted line). \emph{Right:} fiducial and observed CygNW.}
    \label{fig:puff-up}
\end{figure*}

Both effects can be seen in Figure~\ref{fig:puff-up} where we show the $X_E=0.50$ 
simulation $230$~Myr after starting the simulation run, which is the last snapshot
before the central temperature of fiducial CygA stabilises in this simulation. 
In this figure we can also see an increased temperature around $1-2$~Mpc. The
interstitial region between fiducial CygA and CygNW sits at these radii where the
ICM is heated as a result of the merger. Finally we would like to draw attention to
the green dotted vertical line showing the kernel scale radius. The hot core of 
CygNW lies well within the kernel, meaning that this region is represented by too
few SPH particles to be well resolved numerically. As a result, the high-temperature
core of CygNW is washed out in our simulations and the temperature structure will 
be too cold in comparison to the observations later on.

\subsubsection*{Selecting suitable simulation snapshots}
The measured core separation further constrains our simulations of the Cygnus
cluster, in addition to the (already) matching radial profiles. We compute the
core separation for each snapshot of all three initial velocities, and resimulate 
those closest to a projected distance of $\sim$765~kpc for the six inclination
angles. The finer interpolation writes snapshots at a forty times higher rate
(i.e. every 0.25~Myr) to obtain a fiducial core separation as close to the observation
as possible. Note that we adopt the distance between CygA and CygNW as seen in
our \satellite{Chandra} observation rather than using the (much smaller) value
found in the optical dynamical model. We do so because i) the optical core 
separation is obtained from the galaxy surface density peaks, but the optical 
galaxy distribution is sparse, ii) reproducing the optical surface density
showed us that the level of smoothing the galaxy distribution strongly affects
the peak location, and iii) the simulated X-ray surface brightness is compared 
to the observed X-ray surface brightness.
% $701.3$'' = $765 h_{70}^{-1}$~kpc.

A degeneracy exists between the initial (merger) velocity and the inclination angle.
We can use the amount of the merger heating to select which one of the snapshots 
best-represents the observed cluster state. As seen in Figure~\ref{fig:puff-up} in the 
$1-2$~Mpc region, the merger will heat the ICM in-between fiducial CygA and CygNW.
If the clusters approach one another with a higher velocity, the cluster gas will
be prone to more heating than for lower-velocity simulations. At the same time,
increasing inclination angles correspond to a higher physical separation, thus,
the cluster gas will have undergone less heating and the merger-induced heating
will be less profound. 

\subsection{X-ray surface brightness and temperature}
\begin{figure*}
    \includegraphics[width=\textwidth]{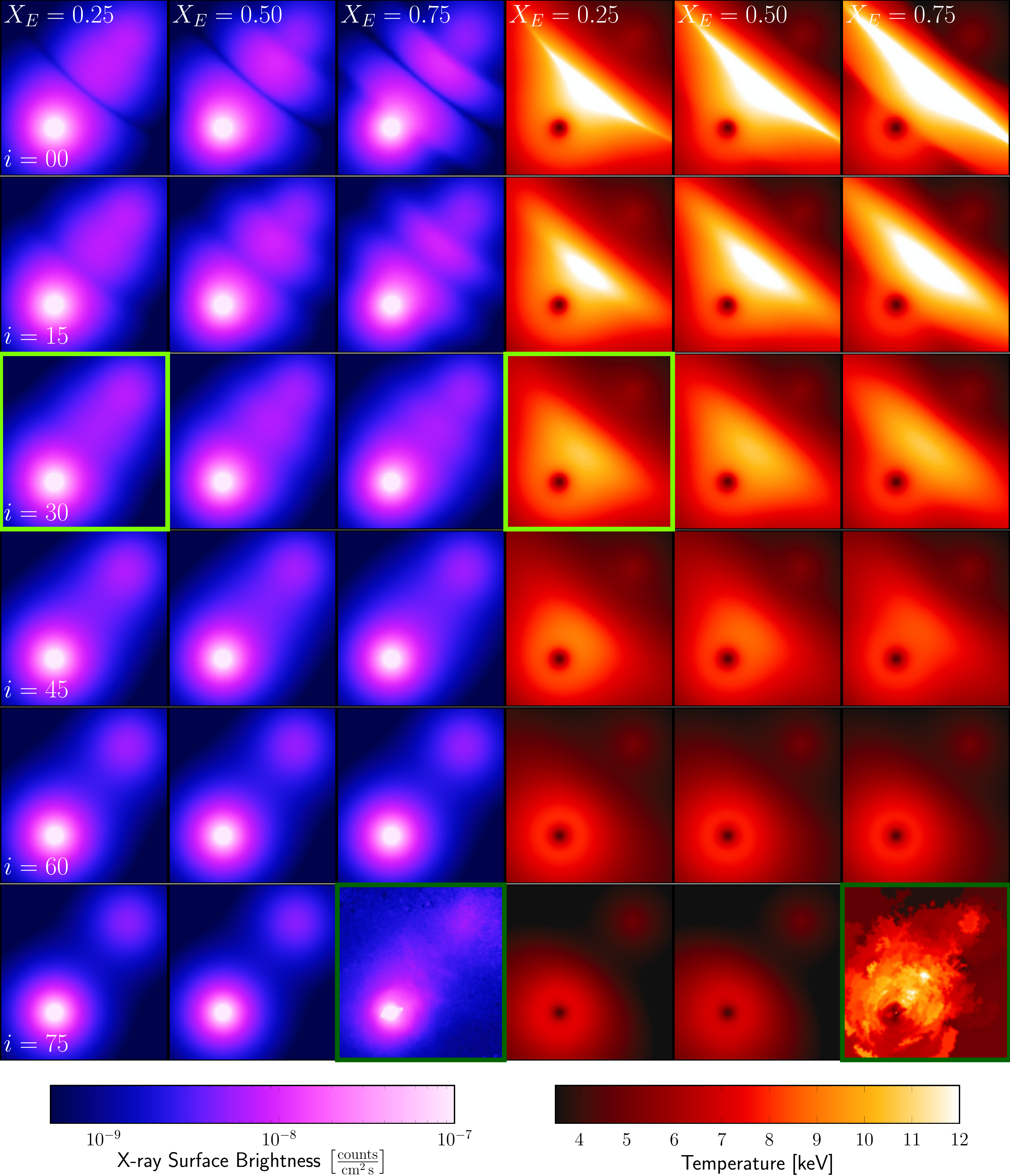}
    \caption{Initial velocity (parametrized by $X_E$) versus inclination angle $i$ in degrees.
            \emph{Left}: X-ray Surface Brightness, \emph{Right}: Spectroscopic
             Temperature. The projected core separation in all panels is $765 \pm 15$~kpc.
             The light green panels show our best fit model, and the observation
             is framed in dark green at the lower right for comparison. }
    \label{fig:XE_vs_EA2}
\end{figure*}

Figure~\ref{fig:XE_vs_EA2} shows the combinations of different values for the
inclination and initial velocity as adopted in this study. 
% For some reason LaTeX gives a fatal error rendering without the pagebreak...
\pagebreak
The selected snapshots are post-processed with 
\code{P-Smac2}\footnote{\href{https://github.com/jdonnert/Smac2}{https://github.com/jdonnert/Smac2}}
\citep{2014MNRAS.443.3564D} to generate two 
dimensional line-of-sight integrated X-ray surface brightness and temperature maps. The
former shows a computation of the projected thermal bremsstrahlung continuum emission, 
adopting \citet[eq. 3]{1996MNRAS.283..431B} and selecting an energy range of $0.5-7.0$~keV
to match the \satellite{Chandra} energy band used to create the observed mosaic.
The latter shows our computations of the projected spectroscopic temperature where 
we use the model of \citet{2004MNRAS.354...10M}. The X-ray surface brightness can 
then be compared to the left panel of Figure~\ref{fig:lss}, and the temperature to 
the right panel. Based on comparison by eye we can already exclude inclination angles
smaller than $30\deg$~and larger than $45\deg$~because the resulting merger-induced 
heating is higher, respectively, lower than observed. We further investigate which
simulation snapshot could best-represent the Cygnus cluster by generating residual maps 
for every combination of the zero-energy orbit fraction $X_E$ and the inclination $i$.
These two-dimensional residuals are then folded into a one-dimensional residual 
histogram to quantitatively identify the `best fit' snapshot. The combination of
$X_E = 0.25$ and $i=30-45$ seems to fits the Cygnus cluster best as their histograms
are centered closest to zero and have the smallest width. However, we are unable to 
accurately distinguish between the three different initial velocities.

\subsection{Best fit simulation snapshot}
\begin{figure*}
    \centering
    \begin{minipage}{.49\textwidth}
        \includegraphics[width=\columnwidth]{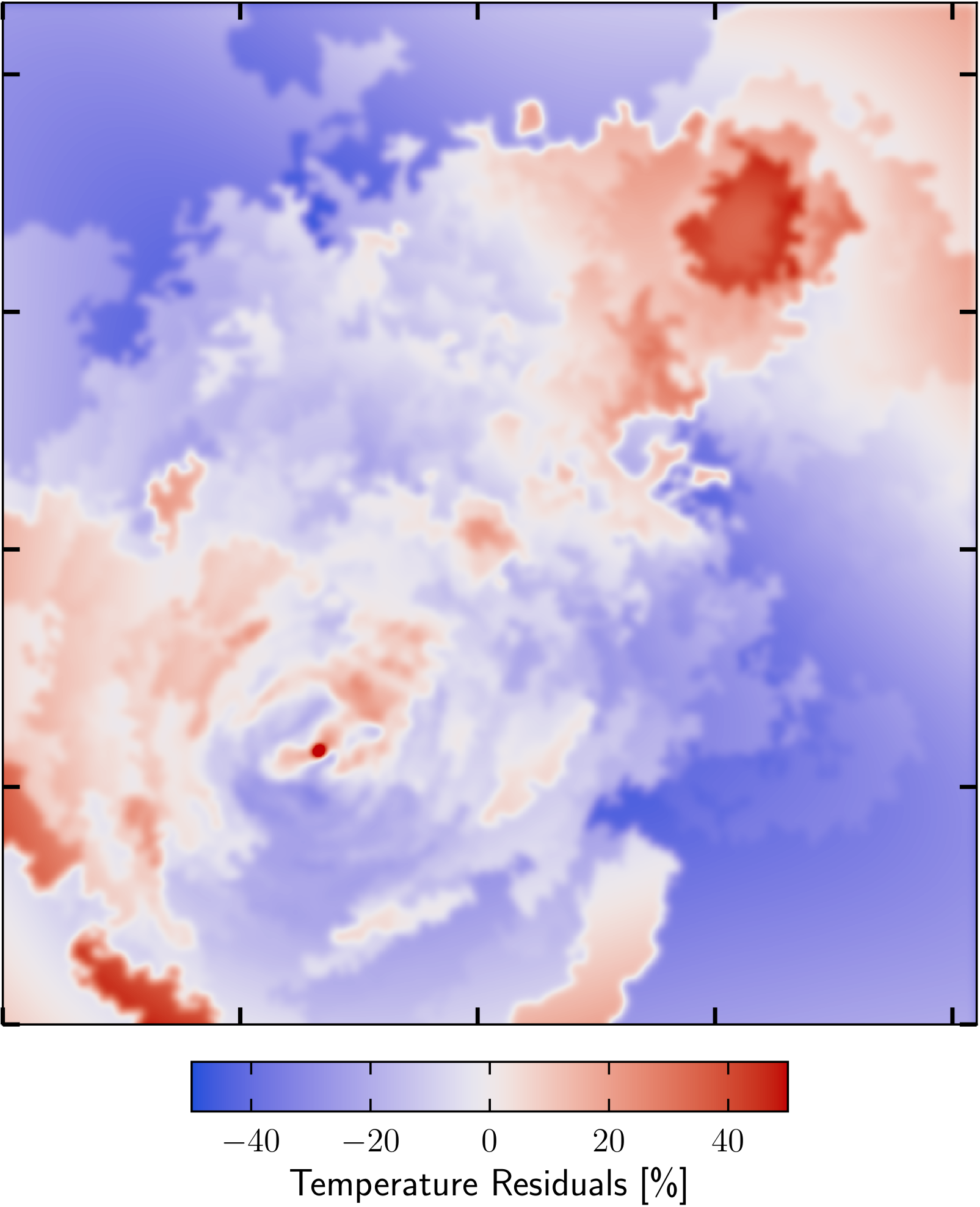}
    \end{minipage}
    \begin{minipage}{.49\textwidth}
        \includegraphics[width=\columnwidth]{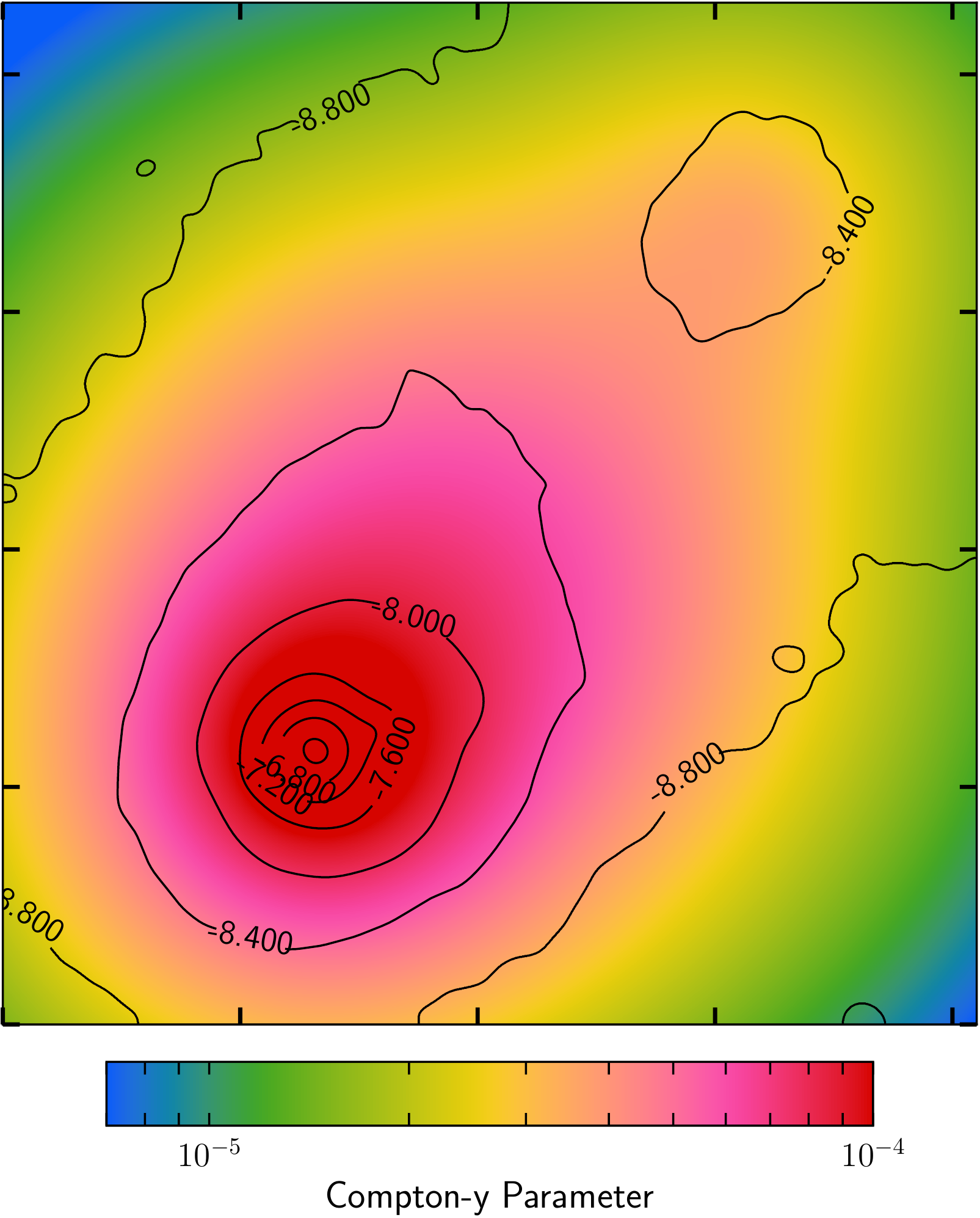}
    \end{minipage}
    \caption{Our best fit $X_E = 0.25$ and $i=30\deg$~simulation. \emph{Left}: 
             Temperature residuals. Red indicates the observed temperature is
             higher than the simulation, conversely, blue indicates a lower
             observed temperature. \emph{Right}: Compton-y parameter.
             In contours the X-ray surface brightness in log (counts/cm$^2$/s) as 
             observed by \satellite{Chandra}.}
    \label{fig:bestmodel}
\end{figure*}

We show a map of residuals for $X_E = 0.25$ and $i=30$ in Figure~\ref{fig:bestmodel}
(left). This particular snapshot has a projected core separation of $765.22$~kpc at
T$=1.59$~Gyr after starting the simulation, and is $\sim270$~Myr prior to 
core passage. We zoom into the simulation box to ensure the physical dimensions
are equal to the box size of Figure~\ref{fig:lss}, and ensure that the centroids of 
CygA and fiducial CygA are located at the same point in the image. We then 
subtract the simulated temperature from the observation, divide by the simulation
and multiply by one hundred. 

Blue colors denote locations of lower simulated temperature, while red indicates a
higher observed temperature. The red dot indicates the location of CygA, and the red
blob in the upper right (north west) shows that CygNW is hotter in the observations 
than in the simulations. We find excess temperature structure in the observations 
that is not due to the hydrostatic temperature structure of the ICM. For example,
we see an arc-like structure (the dark red bin in the bottom left corner). Note, however, 
that this region is underexposed as most of the \satellite{Chandra} pointings are on 
the region directly surrounding CygA. Furthermore, we find excess temperature towards the 
left of CygA, as well as along the presumed merger axis. We further investigate the 
interstitial region between CygA and CygNW to look for indications of a merger shock 
at the interface between both clusters in Section~\ref{sec:shock-or-compression}.

\begin{figure*}
    \includegraphics[width=.47\textwidth]{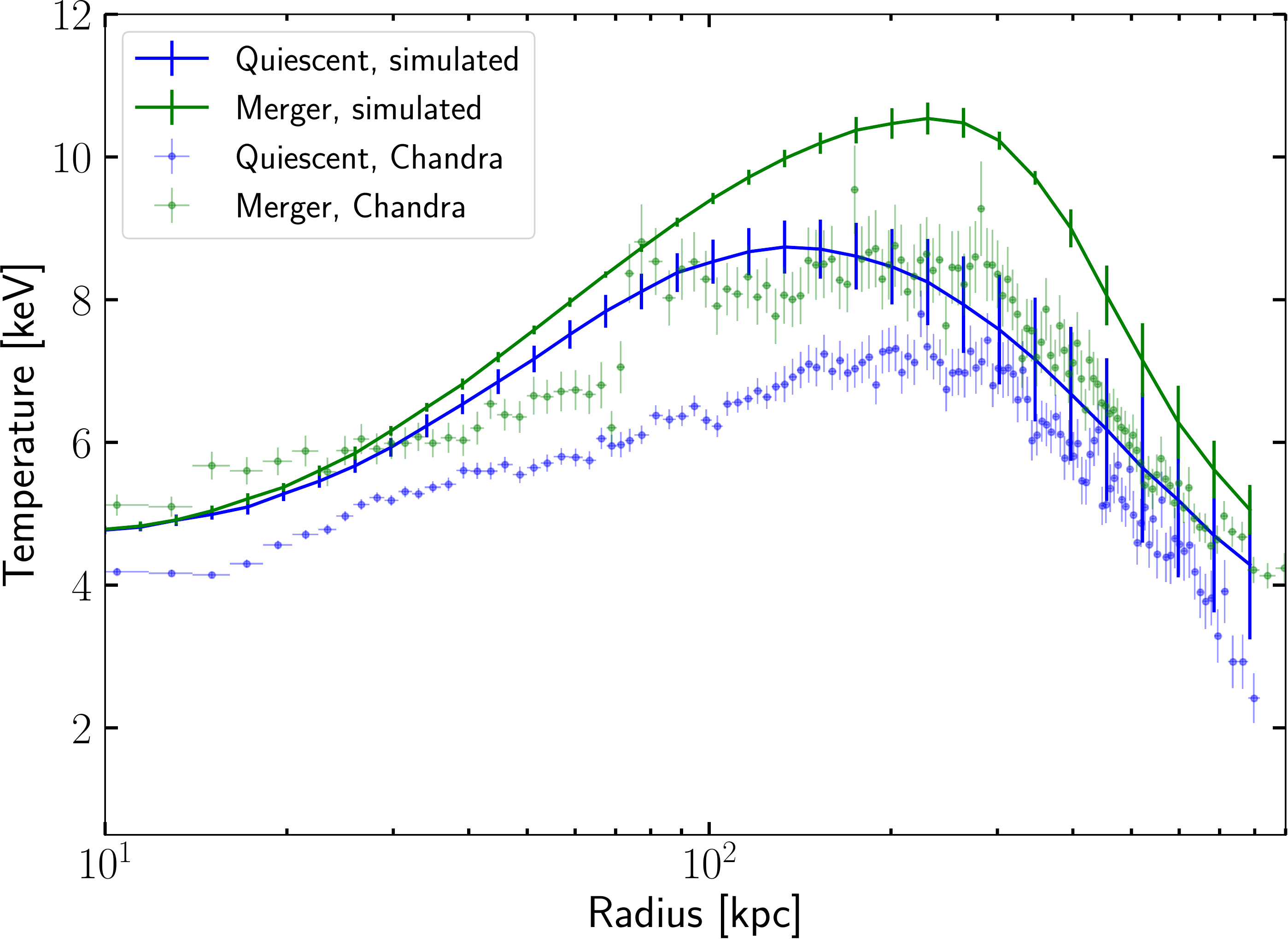}
    \includegraphics[width=.47\textwidth]{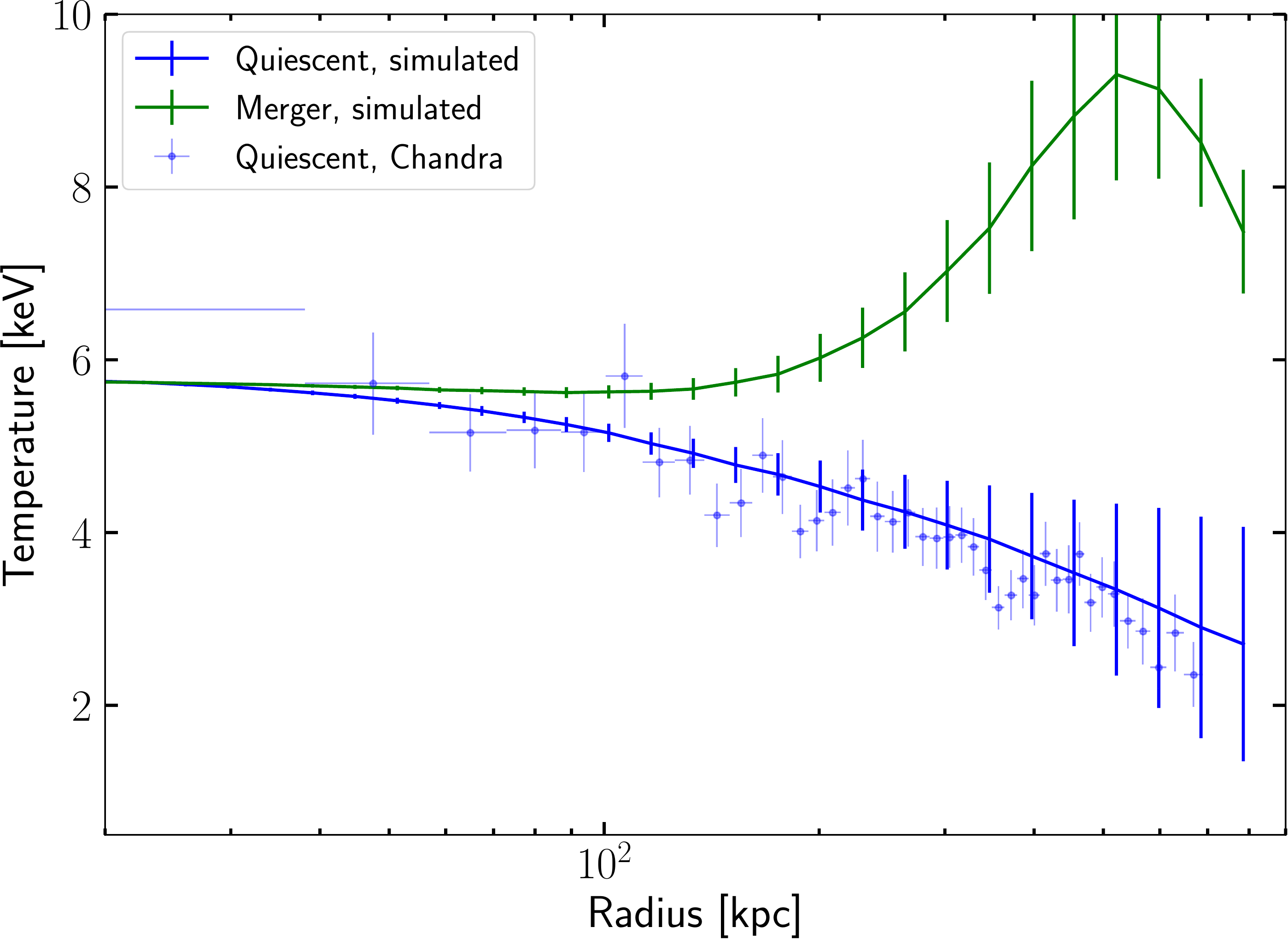}
    \caption{Temperature profiles of the $X_E = 0.25$, $i=30\deg$~simulation after
             $1.59$~Myr with a core separation of $765.22$~kpc.
             The `quiescent' wedge is shown in blue while the `merger' wedge is 
             depicted in green. \emph{Left:} fiducial CygA overplotted on the quiescent
             temperature profile of CygA. \emph{Right:} fiducial and observed CygNW.}
    \label{fig:radial}
\end{figure*}

Furthermore, we extract radial profiles for the haloes that represent CygA and 
CygNW in the best fit simulation snapshot. We use wedges of equal boundary angles
as used for the observation. This provides us with the quiescent and merger radial
profiles (the latter for CygA only) as shown in Figure~\ref{fig:radial}.

\subsection{SZ Compton-y Map}
We compute the expected \citet{1972CoASP...4..173S, 1980ARA&A..18..537S} Compton-y 
parameter, shown in Figure~\ref{fig:bestmodel} (right). Here we overplot the logarithm
of the observed \satellite{Chandra} X-ray contours. Cygnus~A is the brightest
extragalactic radio source in the local Universe so a rather high dynamic range 
will be needed to observe any other radio emission in the vicinity of Cygnus~A that is not
related to the FR-II source. At present no independent Compton-y measurement of the Cygnus 
cluster is available in the literature, to our knowledge.

\subsection{Shock \& compression}
\label{sec:shock-or-compression}
We investigate the interface region between both merging cluster haloes to look for
signs of merger shocks and associated merger-induced shock heating during the 
approach prior to core passage. We use the built-in shock finder 
\citep{2016MNRAS.458.2080B} to obtain Mach numbers and shock speeds for the
SPH~particles. The shock finder picks up less than several hundred 
shocked particles after numerical artefacts from the initial conditions have
settled (at T $\gtrsim 1$~Gyr) but prior to core passage (T $< 1.88$~Gyr). We 
find significant numbers of shocked SPH~particles (10$^{4-5}$) after core passage.

\begin{figure*}
    \includegraphics[width=\textwidth]{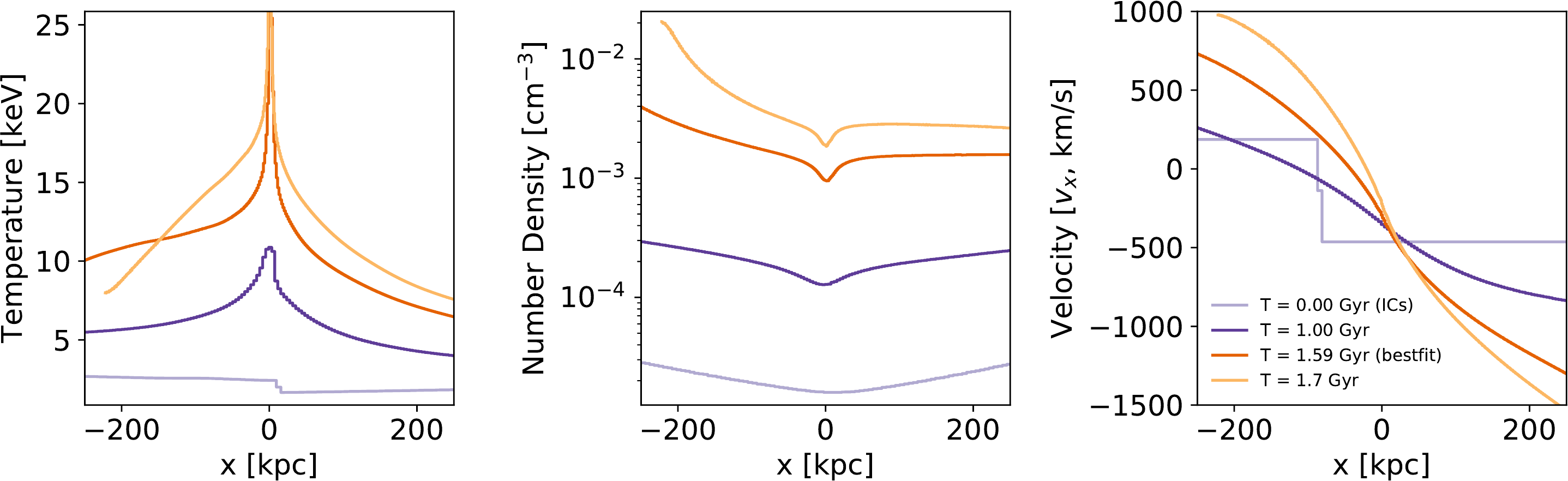}
    \caption{Temperature, density and velocity of SPH particles inside a cylinder of
             radius 150~kpc located along the merger axis and centered at the interface
             between both haloes.}
    \label{fig:noshock}
\end{figure*}

Furthermore, we check whether the Rankine-Hugoniot jump conditions are fulfilled 
for fluid elements located within a cylinder of radius $150$~kpc placed along 
the merger axis (centered on the interface between both merging haloes). 
In Fig.~\ref{fig:noshock} we show the mean temperature, density and velocity 
within this cylinder at four different time steps. We observe no discontinuity 
in density, temperature or velocity in the boundary region between the two clusters. 

\begin{figure*}
    \includegraphics[width=.47\textwidth]{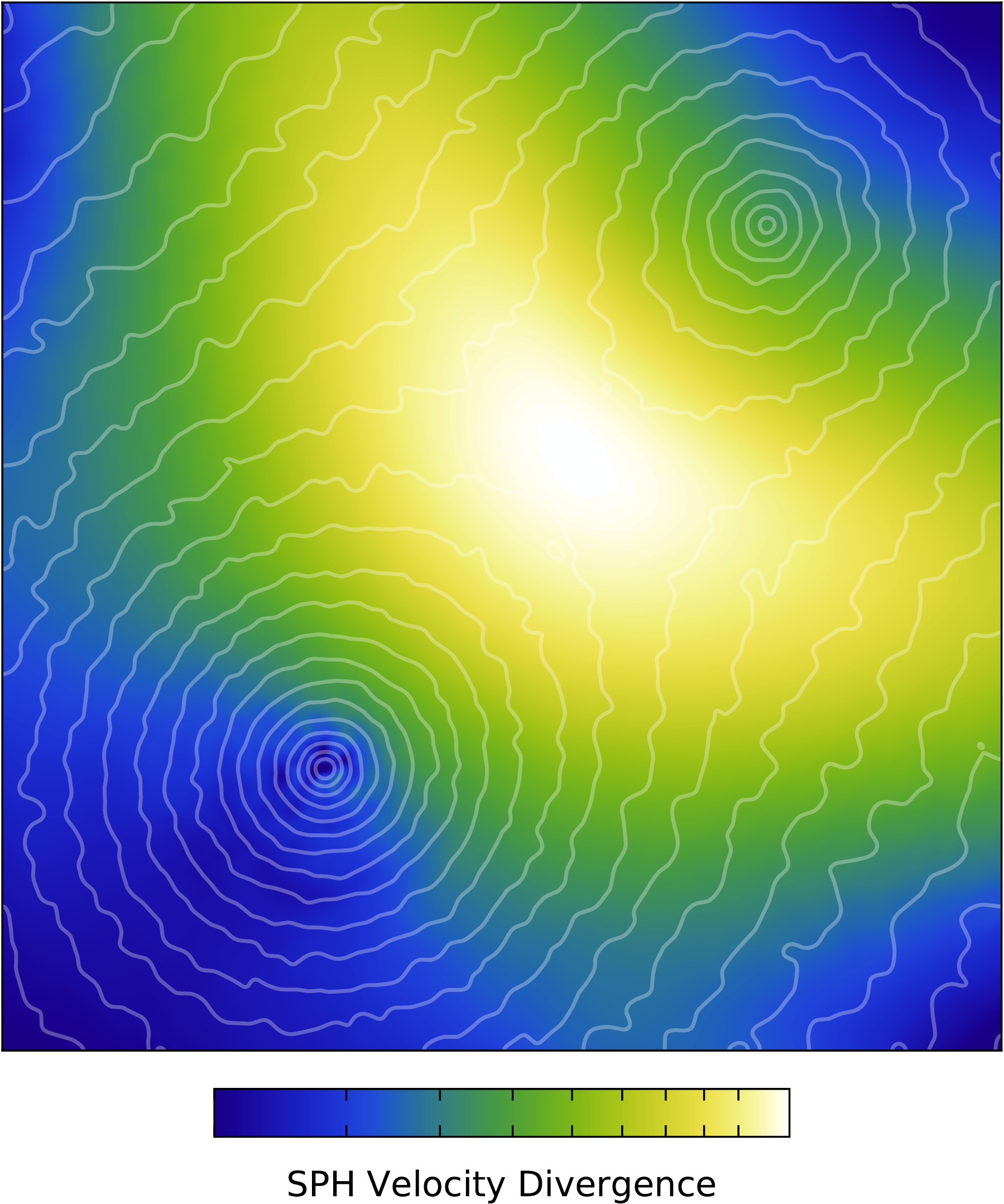}
    \includegraphics[width=.47\textwidth]{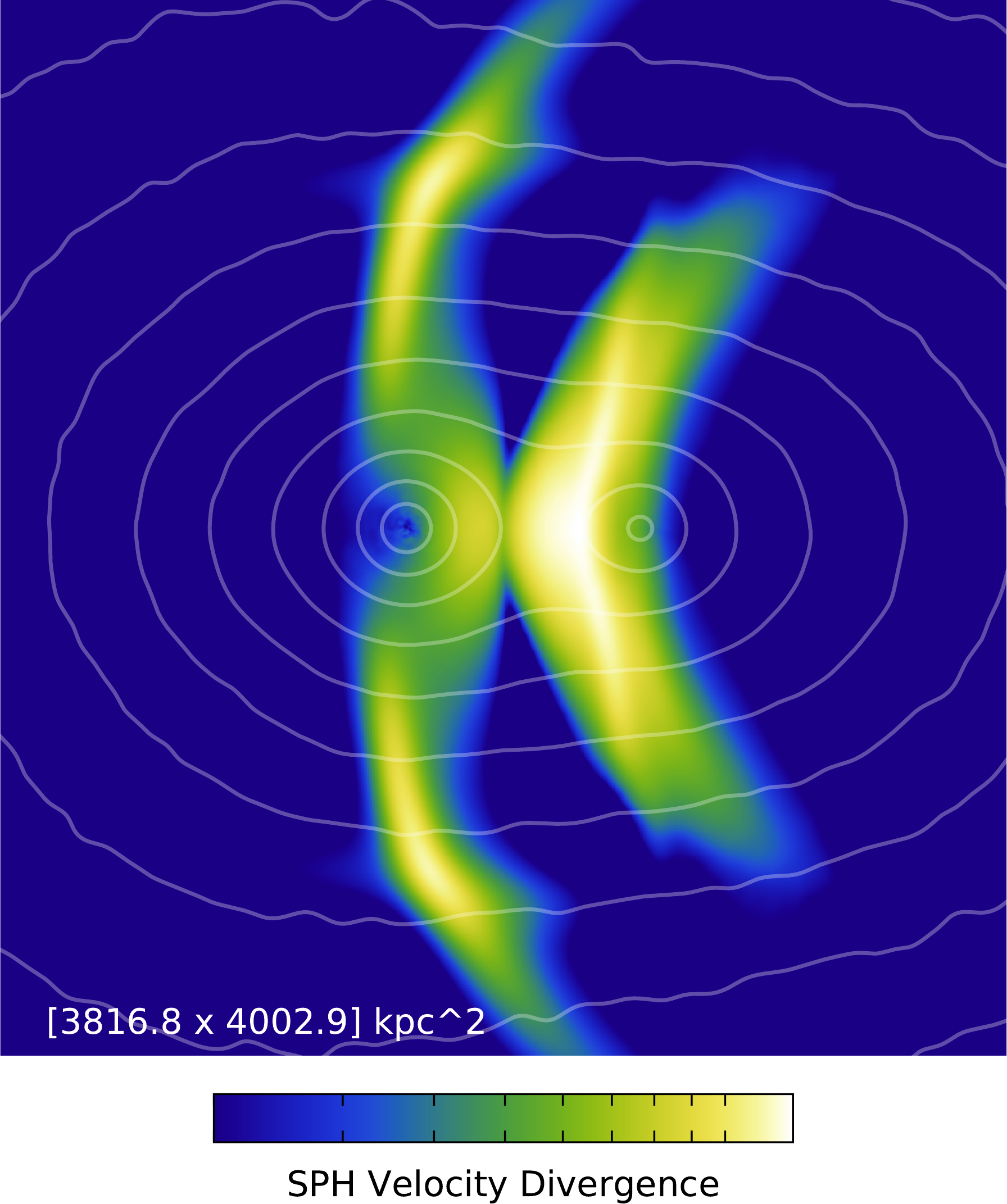}
    \caption{SPH velocity divergence (arbitrary units) for the bestfit snapshot of
             the $X_E=0.25$ and $i=30\deg$~simulation. The sign of the velocity divergence is 
             negative, indicating where compression in the interstitial region occurs
             and is strongest (white). Contour lines show the dark matter density and are
             included to indicate where both haloes can be found (levels differ between both panels).
             To emphasize, both panels are generated using the same simulation snapshot, but viewed
             from different angles. \emph{Left:} Projected ($i=30\deg$), and rotated ($51\deg$)
             map. \emph{Right:} De-projected ($i=0\deg$) picture where the merger-axis is
             horizontal (i.e. the simulation box is rotated in the $xy$-plane with respect to
             the left panel with angle $-51\deg$ measured from west to the north).
             % The physical core-separation here is $862.65$~kpc, close to the required
             % $765 \, h_{70}^{-1}$ / cos($30\deg$) = $883.35$~kpc.
             }
    \label{fig:divv}
\end{figure*}

Finally, we show the SPH velocity divergence (in arbitrary units) of the bestfit 
simulation snapshot ($X_E=0.25$ at T=$1.59$~Gyr) in Fig.~\ref{fig:divv}. The sign 
is negative everywhere within this image and white [blue] indicates where compression
is strongest [weakest].
The panel on the left shows a projected picture for the bestfit $i=30\deg$ inclination
angle. The panel on the right shows the same simulation snapshot, but de-projected 
($i=0\deg$), and de-rotated within the $xy$-plane (with angle $−51\deg$ measured
from west to the north).

\section{Discussion}
\label{sec:discussion}
The baryonic profiles of CygA and CygNW are constrained by two megaseconds of 
\satellite{Chandra} observations of the ICM, and we derive the dark matter 
profile under the assumption of hydrostatic equilibrium and spherical symmetry. 

\subsection{Concentration parameter}
\label{sec:cnfw}
The \satellite{Chandra} observation span a radius range of $50$-$100$~kpc
for the cluster haloes of both CygA and CygNW. Modelling the cluster environment gives virial
radii of ${>}1500$~kpc. This is well outside of the observed data range 
and indicates that the virial radii of CygA and CygNW overlap significantly as the
observed core separation is $701.3'' \sim (765 \, h_{70}^{-1}$~kpc). This makes
it difficult to infer the scaling radius $r_s$ (the turnover point in the NFW 
slope from $-1$ to $-3$), and influences the inferred concentration parameter.
Moreover, this implies that the progenitor clusters have a total gravitating mass 
$M_{200}$ at the virial radius such that the mass-ratio is of order $1.5:1$.
Note, however, that we assume spherical symmetry for both cluster haloes. A more 
realistic scenario is that the cluster haloes are triaxial and elongated along
the cosmic filament, particularly in the scenario of a major merger. The spatial
baryon distribution in CygNW might hint to elongation towards CygA. Assuming
spherical symmetry could result in an uncertainty of the mass-estimate by up to 
a factor two \citep[e.g.][]{2007MNRAS.380..149C} and could change the true
mass ratio significantly.

\subsection{Cut-off radii}
\label{sec:cutoff}
The inferred cluster parameters are furthermore influenced by the cut-off 
in the density profiles. It is particularly difficult to constrain the cut-off
radius. We find values that lie well outside the domain of our observations. Even 
if the cut-off radius would lie within $1$~Mpc, low-photon counts 
and a large surface area in the outskirts of the cluster make for low-number
statistics or an infeasibly large observation time. In addition, this would require
correcting for non-trivial systematic background errors. D17 reports a cut-off radius 
of $1.7$ times the virial radius $R_{200}$ for the Perseus cluster, while our best 
fit for the Cygnus cluster shows a cut-off smaller than the virial radius. Cutting 
the density results in a slightly lower baryon fraction and concentration parameter 
and a higher total mass than a model without an additional cut-off in the density profiles.
% , see Table~\ref{tab:ics} for the values with and without the
% additional cut-off in the density profile.

\subsection{Increasing central temperature}
\label{sec:puffup}
Given our best fit cluster parameters we set up initial conditions that faithfully
represent the observed cluster state in the beginning of the simulation. When 
the haloes are simulated separately in a box with periodic boundaries, we find
that the profiles remain very stable for over $5$~Gyr. When both clusters are 
placed on a merger trajectory in the same box, however, a rapid increase
in the central temperature is noticed during the first $\sim 200$ Myr. We 
interpret this as the result of tidal interaction between both halos. We note that 
an equilibrium with a slightly higher central temperature is reached that remains 
close to the observed \satellite{Chandra} profiles. In addition, we notice that the scatter
of the simulated temperature around the analytical profile increases over the course
of the simulation run. We interpret this behaviour as an artefact of the particular
SPH formulation used: numerical imprecision mostly affects the internal energy (from 
which the temperature is computed) and accumulates over the course of the \code{Gadget3} run. 

\subsection{Best fit simulation snapshot}
\label{sec:bestfit}
We investigate the effect of increasing the initial velocity, parametrized by
the zero-energy orbit fraction $X_E$, as well as changing the orientation of 
the system on the sky by varying the inclination $i$, shown in 
Figure~\ref{fig:XE_vs_EA2}. The inclination is defined as zero when the merger
occurs in the plane of the sky while a value of $90\deg$~corresponds to a merger
along the line of sight. As two distinct cluster haloes are visible the latter is
ruled out. Prior literature values
are $54\deg$~\citep{2013AN....334..346S} or $30-45\deg$~\citep{2005AJ....130...47L}
from dynamical modeling. The true physical distance increases with inclination,
thus simulated time decreases. Therefore the amount of merger heating will be lower
with increasing inclination. We use this to resolve the time-angle degeneracy. 
Inclination angles outside the range $30-45\deg$~can certainly be ruled out because the
amount of merger-induced heating is either much larger, or non-existent for smaller 
or larger inclination angles, respectively.

It is, however, very difficult to obtain a more precise value for the inclination 
angle or to differentiate between different merger velocities. We show radial
temperature profiles in Fig.~\ref{fig:radial}, and two-dimensional residuals map as well as 
the expected SZ signal for the $X_E = 0.25$ and $i = 30$\deg~simulation in
Fig.~\ref{fig:bestmodel}. Observing a distinct merger 
shock and quantifying it with an estimate of the Mach number would allow further
constraints on the merger velocity and inclination angle. Furthermore, radially averaging
the observed temperature profile could wash out the temperature structure at 
larger radii because the bins span a larger area where part of the merger-enhanced
temperature structure is averaged with lower-temperature gas that has not been
affected as much by the ongoing merger. 

We now turn to the residuals map of the temperature structure in Figure~\ref{fig:bestmodel}
(left). The remaining temperature structure is neither the hydrostatic temperature
of the intracluster medium, nor merger-heated gas. CygNW appears hotter in the
observations, but the innermost data point falls below the resolution limit
of our simulations. Therefore the temperature in the innermost region of CygNW is not
well represented in our simulations. The residuals surrounding CygA, on the other
hand, are certainly resolved numerically and show residual temperature structure.
These radii correspond to the temperature jumps in the radial temperature profile of
Figure~\ref{fig:radial} (left). Although the temperature profile in the inner region
`creeps up', the overall radial temperature profile of the `quiescent' and
`merger-enhanced' region are much smoother in the simulation than in the observation.
We suggest that the excess heating could be due to AGN activity over the past hundreds 
of mega years, and could be turned quantitatively to an estimate of the energy output of 
Cygnus~A. Assuming that the AGN outflow travels outward with the local speed of 
sound, the distance can be used as a proxy for time of the outburst. See 
Wise~et~al.~(in~prep).

\subsection{Shocks \& compression}
\label{sec:compression}

We see no indications for a merger shock in the Cygnus cluster in Fig.~\ref{fig:noshock}.
The density, temperature, and velocity increase smoothly over time, except for
artificial discontinuities in the initial conditions that settle within the first
Gyr. The presence of a shock would require that the fluid satisfies the Rankine-Hugoniot
jump conditions. However, we observe no such discontinuities. Furthermore, we notice 
that the velocity changes sign at the interface region $x=0$. This suggests that 
the merger is not yet driving shocks into the ICM, and that the merger-induced 
heating of the ICM as seen in Fig.~\ref{fig:XE_vs_EA2} (three columns on the right)
is the result of compression. Therefore we show the SPH velocity divergence 
in Fig.~\ref{fig:divv} to further illustrate where compression occurs and is
strongest. The location of the largest magnitude of the velocity field (white
colours) correlates exactly with the region of highest merger-enhanced temperature
in the ICM.

% The velocity in the initial conditions is bimodal due to the idealised setup as discussed
% in Section~\ref{sec:orbit}. We note that the initial velocity is slightly offset from the
% interface at $x=0$ because the virial radii of both haloes is used to assign which particles
% receive a positive c.q. negative value, while we assign particles to CygA/CygNW based on
% $x$-values left or right of the center of mass. 

In addition, the non-detection of shocked SPH particles in the interface region 
between $1~\lesssim~T~<~1.88$~Gyr further strengthens the view that we do not see
a merger shock in-between CygA and CygNW. Within this time frame only several hundred 
SPH-particles are picked up by the shock finder, which is significantly less than 
the number of neighbours required to sample the SPH kernel properly. We do not find 
a merger shock in-between both cluster centroids regardless of the exact inclination 
angle of the merger. The observed (projected) core separation of $765 \, h_{70}^{-1}$~kpc
in the $X_E~=~0.25$ simulation is reached at times between T~=~$1.31~-~1.63$~Gyr
for inclination angles $60-0$ degrees, which falls within the time range where we 
find only several hundred shocked particles.

A strong axial shock does develop ahead of the dark matter core of the halo that
represents CygA immediately after core passage, which occurs at T~=~$1.88$~Gyr
for the $X_E=0.25$ simulation. This shock is numerically very well resolved by 
at least 12.000 SPH particles (when the shock first forms), up to well over 
100.000 particles at later times. However, the post core-passage scenario is of
no physical interest to the pre-merger Cygnus cluster. 

The development of merger shocks in our simulations is consistent with the
recent study involving cosmological simulations of \citet{2018ApJ...857...26H},
where the authors studied the development of merger shocks in (near) head-on
merger simulations of mass ratio $~2:1$.  However, we find no equatorial shocks as
a result of our initial conditions setup: we do not correctly sample the background 
density of the cosmic filament in the simulation box, and our simulations start
after or at the time these shocks would (start to) form.

\section{Conclusions}
\label{sec:conclusions}
We ran state of the art (idealised) binary merger simulations of the ongoing merger
and draw the following conclusions. 

\begin{itemize}
    \item{CygA is surrounded by a dark matter halo that is more massive and more 
          concentrated than expected for a typical cluster of galaxies. The inferred
          concentration parameter is a significant outlier in the 
          \citet{2008MNRAS.390L..64D} scatter relation of $c_{\text{NFW}}$ versus 
          $M_{200}$. The estimated baryon fraction is well below ten per cent
          (where we expect a total baryon fraction of seventeen per cent for a 
          typical cluster). The exact value, however, strongly depends on the 
          particular value inferred for the cut-off radius.}
    \item{We interpret the low fraction of baryons in the intracluster medium as an
          indication that a higher fraction of baryons has cooled sufficiently to settle
          into the cluster galaxies, particularly the brightest cluster galaxy. This may
          have resulted in a higher star formation rate and/or more baryons present in the 
          interstellar medium of the Cygnus~A galaxy and could explain the persistent
          high output power of the active galactic nucleus as more fuel is available
          in the Cygnus~A galaxy than in typical brightest cluster galaxies.}
    \item{Modeling the data suggests that the CygNW subcluster has a total 
          gravitating mass at the virial radius that is roughly equal to the total
          mass in CygA. CygNW appears much fainter in the X-ray surface
          brightness than CygA, so we would intuitively expect CygA to be much
          more massive. The X-ray surface brightness, however, scales with 
          $\sqrt{T} \cdot \rho_0^2$, where the central density $\rho_0$ is an 
          order of magnitude lower for CygNW than for CygA. The high temperature 
          of CygNW, on the other hand, could be indicative of the high mass as the 
          initial conversion from potential to thermal energy depends on the mass.} 
    \item{Our simulations show a best fit snapshot with an inclination of $30-45$\deg,
          consistent with the dynamical model of \citet{2005AJ....130...47L}.
          However, the X-ray observations show a core separation of
          $765~h_{70}^{-1}$~kpc rather than the value of $400~h_{75}^{-1}$~kpc 
          found there.}
    \item{We find a total gravitating mass of $7.36 \cdot 10^{14}$~M\Sun \, for 
          CygA and $4.99 \cdot 10^{14}$~M\Sun \, for CygNW. The mass of both
          subclusters found by dynamical modeling of $2 \cdot 10^{15}$~M\Sun \,
          is an order of magnitude higher than found using the X-ray observations.
          Numerical simulations with such a high cluster mass would be inconsistent
          with the \satellite{Chandra} observations.}
    \item{We find no signs of a merger shock in-between CygA and CygNW and suggest
          that the merger heats the ICM due to compression. We subtract the simulated
          temperature structure of the Cygnus cluster from the observations to 
          generate a two-dimensional map of residual temperature structure that
          is neither due to the hydrostatic temperature structure nor due to
          merger-induced heating of the gas. }
\end{itemize}

\section*{Acknowledgements}
We would like to thank the anonymous referee for their useful comments on the
draft version of this paper. The scientific results reported in this article are
based on observations made by the Chandra X-ray Observatory. We thank Klaus Dolag
and Volker Springel for access to the proprietary version of the code \code{Gadget3}.
We thank SURFsara (\href{https://www.surfsara.nl}{www.surfsara.nl}) for the support
in using the Lisa Compute Cluster, and the Max Planck Computing and Data Facility
for maintaining the Freya compute cluster. JD was supported by the People Programme
(Marie Sklodowska Curie Actions) of the European Unions Eighth Framework Programme
H2020 under REA grant agreement no 658912, ``Cosmo Plasmas''. TLRH acknowledges
support from the International Max-Planck Research School (IMPRS) on Astrophysics.
TLRH thanks Alex Arth and Ulrich Steinwandel for useful discussions and help with 
resimulating \code{P-Gadget3} snapshots for finer interpolation. Plots were generated
using \code{Matplotlib} \citep{Hunter:2007}, and the two-dimensional figures in this 
paper use the perceptually uniform colour maps of \citet{2015arXiv150903700K}.
This research made use of \code{NumPy} \citep{van2011numpy}, \code{SciPy} 
\citep{jones_scipy_2001}, \code{Astropy} \citep{2013A&A...558A..33A}, and
\code{APLpy}, an open-source plotting package for \code{Python} 
hosted at \url{http://aplpy.github.com}.

%%%%%%%%%%%%%%%%%%%%%%%%%%%%%%%%%%%%%%%%%%%%%%%%%%



\bibliographystyle{mnras}
\bibliography{CygnusAMerger} % if your bibtex file is called CygnusAMerger.bib

% \begin{thebibliography}{99}
% \bibitem[\protect\citeauthoryear{Author}{2012}]{Author2012}
% Author A.~N., 2013, Journal of Improbable Astronomy, 1, 1
% \bibitem[\protect\citeauthoryear{Others}{2013}]{Others2013}
% Others S., 2012, Journal of Interesting Stuff, 17, 198
% \end{thebibliography}

%%%%%%%%%%%%%%%%%%%%%%%%%%%%%%%%%%%%%%%%%%%%%%%%%%

%%%%%%%%%%%%%%%%% APPENDICES %%%%%%%%%%%%%%%%%%%%%

% \appendix
% 
% \section{Radial Profiles for CygA and CygNW}
% \label{sec:ap-full-profiles}

%%%%%%%%%%%%%%%%%%%%%%%%%%%%%%%%%%%%%%%%%%%%%%%%%%

% Don't change these lines
\bsp    % typesetting comment
\label{lastpage}
\end{document}